\PassOptionsToPackage{prologue,dvipsnames}{xcolor}
\documentclass[sigconf]{acmart}

\usepackage[utf8]{inputenc}
\usepackage{hyperref}
\usepackage{graphicx}
\usepackage{enumitem}
\usepackage{natbib}
\usepackage{booktabs}

\usepackage{listings}

\usepackage{multicol}
\usepackage{multirow}
\usepackage{makecell}
\usepackage{caption}
\usepackage{subcaption}
\usepackage[framemethod=TikZ]{mdframed}

\usepackage{url}

\usepackage{float}

\usepackage{minted}
\usepackage{siunitx}

\newcommand{\secref}[1]{Sec.~\ref{#1}}
\newcommand{\figref}[1]{Fig.~\ref{#1}}
\newcommand{\tabref}[1]{Tab.~\ref{#1}}
\newcommand{\lstref}[1]{Listing~\ref{#1}}

\newcommand{\cell}[2]{%
    \texttt{[{#1},{#2}]}%
}

\newcommand{\tile}[1]{%
    \texttt{#1}%
}

\newcommand{\code}[1]{%
    \texttt{#1}%
}

\newlength{\dpcircle}
\newlength{\rcircle}
\newlength{\dcircle}
\newlength{\scaledcircle}
\newcommand{\docircle}[4]{%
  \setlength{\dpcircle}{\dp\strutbox}%
  \setlength{\dcircle}{\dpcircle}%
  \addtolength{\dcircle}{\ht\strutbox}%
  \setlength{\scaledcircle}{1.25\dcircle}%
  \setlength{\rcircle}{0.5\scaledcircle}%
  \setlength{\unitlength}{1sp}%
  \begin{picture}(\number\scaledcircle,0)
    \color{#1}
    \put(\number\rcircle,\number\dpcircle){\circle*{\number\scaledcircle}}
    \color{#2}
    \put(\number\rcircle,\number\dpcircle){\circle{\number\scaledcircle}}
    \put(\number\rcircle,0){\makebox[0pt]{\textcolor{#3}{#4}}}
  \end{picture}%
}

\definecolor{circleyellow}{rgb}{0.99, 0.95, 0.8}
\definecolor{circleyellowline}{rgb}{0.84, 0.71, 0.34}

\newcommand{\colornumber}[1]{%
  {\footnotesize\docircle{circleyellow}{circleyellowline}{black}{\textbf{#1}}}%
}

\newcommand{\step}[1]{%
  Step~\colornumber{#1}%
}

\newcommand{\steps}[2]{%
  Steps~\colornumber{#1}--\colornumber{#2}%
}

\definecolor{circlegreen}{rgb}{0.85, 0.95, 0.85}

\newcommand{\colorgreennumber}[1]{%
  {\footnotesize\docircle{circlegreen}{black}{black}{\textbf{#1}}}%
}

\definecolor{exampletop}{gray}{0.9}
\definecolor{examplebottom}{gray}{0.8}

\def\formtmp#1#2{{\vskip12pt\noindent\fboxsep=0pt\colorbox{#1}{\vbox{\vskip3pt\hbox to \linewidth{\hskip3pt\vbox{\raggedright\noindent\textbf{Example}~#2\vphantom{Qy}}\hfill}\vspace*{3pt}}}\par\vskip2pt%
\noindent\kern0pt}}

\newenvironment{exampleblock}[1]{\ignorespaces\def\stmtopen##1{##1}%
\formtmp{exampletop}{#1}}{\par\noindent\textcolor{examplebottom}{\rule{\columnwidth}{1pt}}\vskip2pt\par\addvspace{\baselineskip}}%

\newcommand{\viatra}{\textsc{Viatra}}

\lstdefinelanguage{Xtend}
{
    keywordstyle=[1]{\color{eclipsekeywordred}\bfseries},
    keywordstyle=[2]{\color{darkblue}},
    keywordstyle=[3]{\color{xtendorange}\textit},
    morekeywords=[1]{val, it, override, return},
    morekeywords=[2]{changeBurningToRoad, startTile, startTileFixedLocation},
    morekeywords=[3]{createRule, agentInTerminatingState, agentInNonTerminatingState, instance},
    morekeywords=[2]{container},
    extendedchars=true,
    literate={container}{{{\textcolor{darkblue}{container}}}}1,
    morekeywords=[2]{policyService},
    extendedchars=true,
    literate={policyService}{{{\textcolor{darkblue}{policyService}}}}2
}
\lstdefinelanguage{VQL}
{
    keywordstyle=[1]{\color{eclipsekeywordred}\bfseries},
    keywordstyle=[2]{\color{darkblue}\bfseries},
    morekeywords=[1]{pattern, or, find, true},
    morekeywords=[2]{Tile, Forest, BurningTile, states, x, y, GoalTile, StartTile},
}
\definecolor{autoblue}{HTML}{cce4ff}
\definecolor{darkblue}{rgb}{0.0,0.0,0.6}
\definecolor{darkred}{rgb}{0.6,0.0,0.0}
\definecolor{domaingreen}{RGB}{209, 233, 143}
\definecolor{eclipsekeywordred}{RGB}{127,0,85}
\definecolor{mauve}{rgb}{0.58,0,0.82}
\definecolor{xtendorange}{RGB}{171,48,0}

\usepackage{eso-pic}
\definecolor{linkblue}{HTML}{006fbd}
\newcommand{\hreffinternal}[3]{\href{#1}{\textcolor{#3}{#2}}}
\newcommand{\hreff}[2]{\hreffinternal{#1}{#2}{linkblue}}

\newcommand{\placetextbox}[3]{
  \setbox0=\hbox{#3}
  \AddToShipoutPictureFG*{
    \put(\LenToUnit{#1\paperwidth},\LenToUnit{#2\paperheight}){\vtop{{\null}\makebox[0pt][c]{#3}}}%
  }%
}%

\setcopyright{acmlicensed}
\copyrightyear{2016}
\acmYear{2026}
\acmDOI{XXXXXXX.XXXXXXX}
\acmConference[MODELS '26]{ACM/IEEE 29th International Conference on Model Driven Engineering Languages and Systems}{October 4--9, 2026}{Malaga, Spain}

\begin{document}

\placetextbox{0.5}{0.99}{\large\colorbox{gray!3}{\textcolor{WildStrawberry}{\textbf{Author pre-print.}}}}%

\placetextbox{0.5}{0.97}{\large\colorbox{gray!3}{\textcolor{WildStrawberry}{Publication accepted for the \hreff{https://conf.researchr.org/home/models-2026}{ACM/IEEE International Conference on Model Driven Engineering Languages and Systems (MODELS'26)}.}}}%

\placetextbox{0.5}{0.05}{\colorbox{gray!3}{\textcolor{WildStrawberry}{Author pre-print. Publication accepted for} \hreff{https://conf.researchr.org/home/models-2026}{MODELS'26}.}}%

\title[A Model-Driven Approach for Developing Families of Reinforcement Learning Environments]{A Model-Driven Approach for\\ Developing Families of Reinforcement Learning Environments}

\author{Xiaoran Liu}
\orcid{0009-0000-9908-7406}
\affiliation{
  \institution{McMaster University}
  \city{Hamilton}
  \country{Canada}
}
\email{liu2706@mcmaster.ca}

\author{Istvan David}
\orcid{0000-0002-4870-8433}
\affiliation{
  \institution{McMaster University}
  \city{Hamilton}
  \country{Canada}
}
\email{istvan.david@mcmaster.ca}

\renewcommand{\shortauthors}{Xiaoran Liu and Istvan David}

\begin{abstract}
Virtual training environments are software-intensive systems in which reinforcement learning (RL) agents learn, adapt, and demonstrate meaningful behavior. Virtual training environments offer a safe and cost-efficient alternative to training agents in real-world settings. However, to converge, most realistic RL problems require training in multiple, mostly similar but slightly different environments---i.e., families of environment variants. The typical development process of environment families is a labor-intensive and error-prone manual endeavor that does not scale well. To alleviate these issues, in this paper, we propose a model-driven approach for developing families of RL training environments.
To obtain the family of environments, we develop an approach and prototype tool. In our approach, a hybrid genetic algorithm---a combination of population-based global search and heuristic local search---generates environment families. Mutations and constraints are expressed as model transformations and are operationalized into a search process by a state-of-the-art model transformation engine.
We demonstrate the soundness of our approach in a wildfire mitigation scenario and curriculum learning---a particular learning paradigm that relies on environment families.
\end{abstract}


\maketitle

\section{Introduction}

Reinforcement learning (RL)~\cite{sutton1998reinforcement} has emerged as a popular machine learning technique recent years~\cite{figueiredoprudencio2024survey}. In RL, an autonomous agent explores the state space, and through trial and error, it learns beneficial actions to solve complex problems. This approach is particularly useful in problems where prior training data is scarce or unavailable, e.g., control problems in robotics~\cite{marcinandrychowicz2020learning} and digital twins~\cite{david2024automated}.
Model-driven engineering (MDE), too, has adopted RL to address various technical problems, such as derivation of in-place model transformations~\cite{eisenberg2021towards} and complex transformation chains~\cite{dagenais2025complex}, automated repair of models~\cite{barriga2022parmorel}, and inference of simulation models~\cite{david2022devs}.

Virtual training environments are software systems in which agents learn, adapt, and demonstrate meaningful behavior~\cite{kim2021survey}. By modeling real-world settings and allowing the agents to interact with this model, virtual training environments offer a safe and cost-efficient alternative to training in real-world settings.
However, agents trained in a single, fixed environment typically fail to generalize beyond their training conditions and tend to memorize environment-specific behaviors rather than learning generalizable strategies~\cite{malik2021when}. \citet{cobbe2020leveraging} demonstrate that diverse environment scenarios are essential to adequately train RL agents.
Most realistic RL problems, therefore, require the development of \textit{families} of sufficiently diverse training environments~\cite{tobin2017domain} to improve key properties of the agent (e.g., faithfulness) or the learning process (e.g., training time); or often, simply just to converge to any solution at all.
Multiple training paradigms leverage such environment families. A pertinent example is curriculum learning~\cite{bengio2009curriculum}, in which agents are trained on progressively harder environments to allow early knowledge to improve learning performance. Such a structured progression helps RL agents learn faster, as well as improving the final performance of the trained model---similar to how humans learn concepts step-by-step. Other examples include multi-task learning~\cite{zhang2022survey}, meta-learning~\cite{hospedales2022meta-learning}, and domain randomization~\cite{wang2025rgdr}.

Recent studies indicate that agents may require exposure to \textit{thousands of environment variants} to achieve meaningful generalization~\cite{cobbe2020leveraging}.
Despite this, developing a set of structurally valid training environments that are sufficiently similar but diverse at the same time, and sequencing these environments into an effective training process is still largely a manual effort~\cite{narvekar2020curriculum}. Apart from being a clearly labor-intensive and error-prone endeavor, manual development simply does not scale with the requirements posed by modern RL applications~\cite{dennis2020emergent}.
Thus, automated engineering methods that alleviate human intervention are much sought-after.

In this paper, we propose a model-driven approach for developing families of RL environments. In our approach, an RL expert develops the initial learning environment from which new variants are generated and together, form the family of environments that is required for the RL training. We use a hybrid genetic algorithm (GA)~\cite{moscato1989evolution}, i.e., a combination of population-based global search to generate candidate families of environments, and heuristic local search to identify the best-fitting family. The GA performs the global search by mutating the initial environment by mutation operators codified as model transformations, and structural constraints codified as graph queries---all operationalized in the state-of-the-art \viatra{} model transformation framework~\cite{bergmann2015viatra}. The search is guided by domain-specific diversity measures across individual environments in a population. The repeated application of mutating model transformations yields a potentially large but tractable number of varied and diverse models. Subsequently, constructing a family of environments is achieved by automatically selecting a subset of these varied models using an appropriate local search method (e.g., simulated annealing).
Eventually, it is the lifting of structural and RL-specific properties to the modeled level that enables intelligent mutation strategies of the initial (and downstream) environments.

We evaluate our approach on a curriculum learning problem.
The results indicate that our approach improves learning performance while requiring minimal human intervention at the beginning of the process. We observe families of environments that are individually insufficient for training, but together, they facilitate effective learning.\footnote{Replication package: \url{https://zenodo.org/records/21347611}.}
Our work underscores the utility of the MDE body of knowledge in machine learning and responds to the calls for further research on MDE solutions for reinforcement learning~\cite{naveed2024model}.
\section{Background}\label{sec:background}

\subsection{Reinforcement learning}\label{sec: RL}

\subsubsection{Formal underpinnings}

Reinforcement learning (RL)~\cite{sutton1998reinforcement} is a machine learning paradigm in which an agent interacts with its environment to learn optimal strategies for sequential decision making. 
RL can be formalized as a Markov decision
process (MDP) $\langle \mathcal{S}, \mathcal{A}, \mathcal{P}, \mathcal{R} \rangle$~\cite{puterman1990markov}.
$\mathcal{S}$ denotes the set of observable states, $\mathcal{A}$ is the set of available actions, $\mathcal{P}: \mathcal{S} \times \mathcal{A} \times \mathcal{S} \rightarrow [0,1]$ is the state transition probability function, where $P(s' \mid s, a)$ denotes the probability of transitioning to state $s'$ after taking action $a$ in state $s$, and $\mathcal{R}: \mathcal{S} \times \mathcal{A} \rightarrow \mathbb{R}$ specifies the reward function. At each time step $t$, the agent observes state $s_t \in \mathcal{S}$, selects action $a_t \in \mathcal{A}$ according to policy $\pi(a|s)$, receives reward $r_t = \mathcal{R}(s_t, a_t)$, and transitions to state $s_{t+1}$. The agent aims to learn policy $\pi^*$ that maximizes the expected cumulative reward.

\subsubsection{Training environments} 
RL agents learn through trial and error, but this does not necessarily have to happen in a real-world environment.
Virtual and cyber-physical~\cite{liu2025ai} RL training environments encapsulate a simulation infrastructure that provides a virtual world for agents to interact with. Such environments typically consist of an environment core that receives actions, updates the environment state, and returns observations and rewards; an underlying simulator that executes the probabilistic mechanisms representing the real-world phenomenon~\cite{ross2022simulation}; and a simulator adapter that translates the instruction set and results of a simulator to the world semantics of the agent~\cite{liu2026reference}.
Training environments are critical in the success of RL, as they directly shape the agent's decision-making and actions~\cite{ntentos2024supporting}.
Some of the widely used RL environments include Gymnasium~\cite{towers2025gymnasium} for single-agent tasks, PettingZoo~\cite{terry2021pettingzoo} for multi-agent settings, and Isaac Gym~\cite{makoviychuk2021isaac} for robot learning.
In this work, we use Gymnasium.

\subsubsection{Curriculum learning}
In this work, we use curriculum learning (CL) as the representative example of learning paradigms that rely on families of environments.
CL is a training strategy in which the learning process is organized as a sequence of training criteria that evolve over time~\cite{wang2022survey}, typically from simpler to more complex settings~\cite{bengio2009curriculum}.
This allows for learning elementary skills in lower-complexity environments faster and using those acquired skills in higher-complexity environments later.
The most common structure for a curriculum is a sequence curriculum, where the agent is trained in environments that progress from simpler to more complex. Curricula may also be represented as a directed acyclic graph (DAG)~\cite{thulasiraman2011graphs}, in which vertices correspond to environments and edges indicate which should be trained before others.

\subsection{Model-driven optimization and genetic algorithms}
Model-driven optimization (MDO) applies principles from MDE to search-based optimization problems by representing candidate solutions as models~\cite{john2019searching,john2023graph}. 
In this paradigm, the search space is defined over structured models that explicitly capture domain constraints and relationships. This makes MDO particularly well-suited for problems where solution validity and structure are central concerns. 
In \emph{model-based} MDO, candidate solutions are the models themselves, and optimization proceeds by directly transforming these models through MTs.
To search this structured space, MDO often employs genetic algorithms (GA), an optimization technique inspired by natural evolution~\cite{eiben2015introduction}. 
In GA, a population of candidate solutions is varied over several generations under selection pressure to produce solutions of increased quality over time. 
In GA-based MDO, the population is initialized with model instances, either randomly or by varying a single seed model \cite{burdusel2021automatic}.
In parent selection, parent models are probabilistically chosen from the population to undergo variation and produce offspring. 
Mutations are defined as MTs over an initial model.
Survivor selection then decides which candidate models will be part of the next generation, which may be done based on evaluation. Evaluation assigns a real-valued measure to each model, often referred to as fitness.
This process continues for several generations until termination.

In this work, we use a hybrid GA that combines population-based global search and heuristic local search.
\begin{figure}[t]
    \begin{subfigure}{0.4\linewidth}
        \centering
        \includegraphics[width=0.9\linewidth]{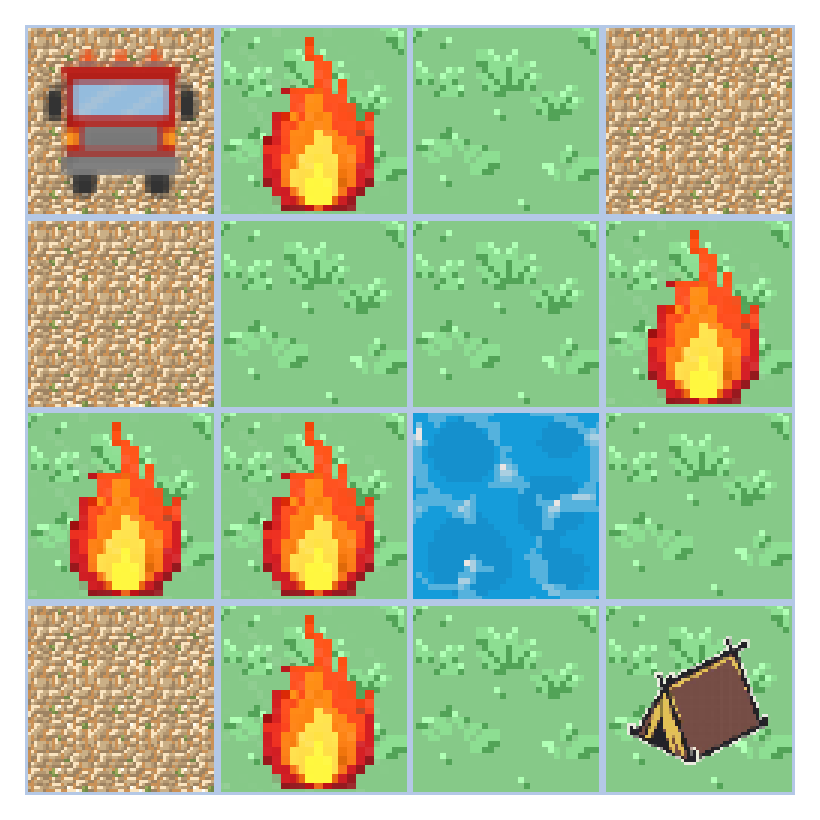}
        \caption{Burning Forest instance}
        \label{fig:running-example-lake}
    \end{subfigure}
    \begin{subfigure}{0.56\linewidth}
        \centering
        \includegraphics[width=1\linewidth]{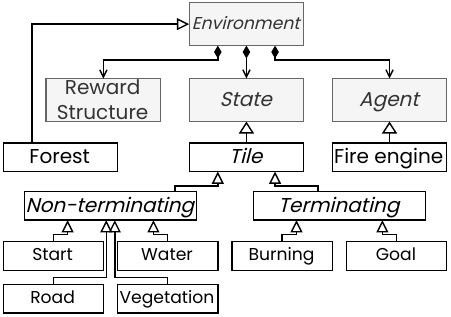}
        \caption{Metamodel}
        \label{fig:running-example-metamodel}
    \end{subfigure}
    \caption{Example \textit{Burning Forest} instance and its metamodel}
    \label{fig:running-example}
\end{figure}

\section{Illustrative example}
\label{sec:running-example}
To illustrate our approach throughout the paper, we draw on a case in wildfire mitigation.
The \textit{Burning Forest} (\figref{fig:running-example}) is a custom RL environment developed using Gymnasium.
It is a topological grid world that acts as the \textit{Environment} on which an \textit{Agent} trains.
It consists of different tiles as its \textit{State}s, including the \textit{Start} (top-left) and \textit{Goal} (bottom-right) tiles, and \textit{Road}, \textit{Vegetation}, \textit{Burning}, and \textit{Water} tiles in between.
The \textit{Agent} navigates from the \textit{Start} to the \textit{Goal} of a rescue mission by selecting one directional action per step (up, down, left, right)  while attempting to avoid burning tiles. Roads are preferred over paths through vegetation as the latter may damage the vehicle. Strategically placed \textit{Water} reservoirs aid rescue missions and thus, should be visited by the agent during its action.

Through trial and error, the agent gradually learns which actions are beneficial in a specific state to reach the goal without stepping on burning tiles.
To guide the learning, the \textit{Reward structure} specifies the reward associated with specific outcomes of chosen actions: $+10$ upon reaching the goal; $-1$ for stepping on a burning tile, $+1$ for reaching a water source for the first time; $0$ for each step on a road tile; and $-0.1$ for each step on a vegetation tile, reflecting the cost of damaging the forest. This is summarized in \tabref{tab:rewards-forest}. 

\begin{table}[t]
    \centering
    \renewcommand{\arraystretch}{0.66}
    \small
    \caption{Burning Forest reward structure}
    \label{tab:rewards-forest}
    \vspace{-1em}
    \begin{tabular}{@{}p{3.2cm}p{3.4cm}r@{}}
    \toprule
         \multicolumn{1}{c}{\textbf{Event}} & \multicolumn{1}{c}{\textbf{Consequence}} & \multicolumn{1}{c}{\textbf{Reward}} \\ \midrule
         Stepping on a road tile & Continue episode & $0$\\
         Stepping on a vegetation tile & Continue episode & $-0.1$\\
         Reaching a water source & One-time reward; continue ep. & $+1$\\
         Encountering a burning tile & Terminate episode & $-1$\\
         Reaching the goal tile & Terminate episode & $+10$\\ \bottomrule
    \end{tabular}
\end{table}

The RL expert trains the agent to navigate the Burning Forest. Training the agent on the actual problem may be too complex for the agent and may render the training inefficient due to delayed convergence to the learning objectives. For example, in \figref{fig:running-example}, the density of burning tiles allows only two successful paths to the goal (through the water reservoir). The agent may need numerous training episodes to discover this path.

To improve learning performance, the RL expert applies \textit{curriculum learning}, i.e., trains the agent on a sequence of progressively more challenging environment variants, starting with an easier problem before exposing it to the hardest one (\figref{fig:running-example-lake}).
\section{Approach}\label{sec:approach}

We use GA to generate families of RL environments. GA is a metaheuristic technique that explores large search spaces by optimizing an objective (e.g., the complexity of environments). GA performs a global search by evolving an initial set of individuals via predefined mutation operators. For model-based generation, mutation operators can be naturally expressed as model transformations to explore the space of potential environment models by guiding the search towards models that fit particular criteria.

\begin{figure*}
    \centering
    \includegraphics[width=0.66\linewidth]{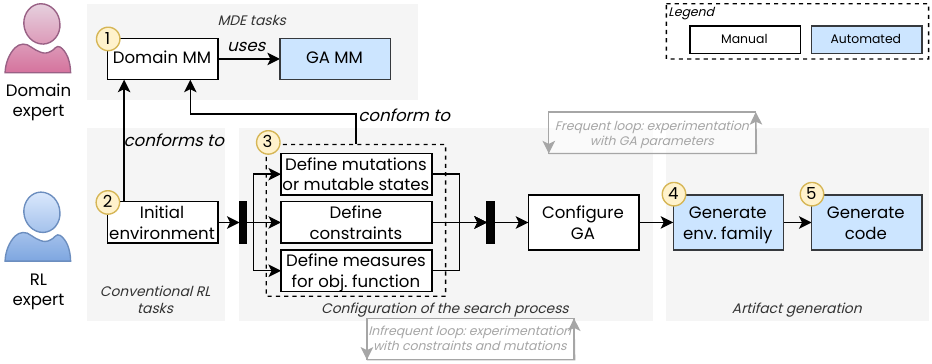}
    \caption{Overview of the approach}
    \label{fig:process}
\end{figure*}

\figref{fig:process} provides an overview of our approach.
\steps{1}{3} are enacted by the \textit{Domain expert} and the \textit{RL expert}.
In \step{1}, the \textit{Domain expert} prepares the activities of the \textit{RL Expert} by defining a suitable \textit{domain metamodel (MM)}.
In \step{2}, an \textit{RL expert}---e.g., RL engineer or software developer---specifies the environment the agent is meant to learn on. This environment is treated as the most challenging one from which less complex variants are to be generated.
In \step{3}, the search process is configured by defining (i) the mutation operators or mutable states for mutation generation; (ii) the constraints for individual environment candidates and the family itself; (iii) the individual- and family-level measures (e.g., diversity of a family of environments, complexity of an individual environment), and (iv) the key GA parameters.

In \step{4}, a GA generates the family of environments through multiple iterations, each of which generates a population of environments with potentially different complexities, optimizing for diversity.
Finally, in \step{5}, the family is sampled in accordance with the training strategy and RL framework-specific code is generated. Subsequently, the training process is ready to be executed. The automation of \step{4} and \step{5} enables RL experts to experiment with various configurations of the GA (``\textit{frequent loop}'') to obtain an environment family that is deemed appropriate. In some cases, more foundational changes may be required, and the ML expert may redefine constraints or mutations (``\textit{infrequent loop}'').

In the following, we elaborate on each of these five steps.

\subsection{Domain modeling and other MDE tasks}

In \step{1}, the domain expert defines the domain metamodel. This metamodel offers a language to express the key characteristics of the RL environment in a subsequent step, as well as to define model transformations. The domain metamodel defines the states of the RL problem that are observable during training. The metamodel also relates these states through an appropriate formal structure, e.g., topological alignment in 2D and 3D problems, and logical alignment in more abstract state spaces, such as design space exploration.

\begin{exampleblock}{Domain (meta)modeling}%
In \figref{fig:running-example-metamodel}, the \textit{Forest} environment and its various states are defined by the domain expert, as well as the agent, which is the \textit{Fire engine} in this particular problem.
\end{exampleblock}

The metamodel is aligned with the GA metamodel. Each environment corresponds to an individual in the GA and is being gradually mutated as the algorithm progresses. (See \secref{sec:approach-ga}.)

In this step, domain-specific languages may be developed to aid the subsequent efforts of the RL expert. Notably, a DSL can help lift the specification of mutations and automate the task of manual transformation specification altogether. Such features are not in the scope of our current work.

\subsection{Defining the initial environment}\label{sec:approach-initial-env}

In \step{2}, the RL expert defines the 
initial environment.
This environment is a fixed element of the eventually generated family of environments, and the starting point for generating the rest of the family.
Different learning paradigms choose initial environments differently. For example, in curriculum learning, the initial environment typically corresponds to the most challenging environment; to the environment without any simulation randomization in domain randomization~\cite{tobin2017domain}; and to a representative task configuration (e.g., a single object with a basic manipulation skill) in multi-task learning~\cite{yu2020meta-world}.
Alternatively, the expert may specify a different starting configuration, e.g., the easiest one, from which variants are generated. 
For example, \citet{florensa2017reverse} construct a curriculum over initial states by progressively moving the starting position from near the goal to increasingly distant locations.
Our approach is able to accommodate any initial environment due to its ability to mutate environments both towards lower and higher complexity.

\begin{exampleblock}{Defining the initial environment}%
In the running example, the environment model defines the \tile{Forest} composed of \tile{Tile}s, specialized into \tile{BurningTile}, \tile{RoadTile}, \tile{VegetationTile}, \tile{WaterTile}, \tile{StartTile}, and \tile{GoalTile}, with a \tile{FireTruck} agent whose \tile{currentState} references a single tile, as shown in \figref{fig:running-example-metamodel}.
The expert specifies the initial environment model as a Burning Forest containing 16 \tile{Tile}s, with a \tile{StartTile} at \cell{0}{0}, a \tile{GoalTile} at \cell{3}{3}, a \tile{WaterTile} at \cell{2}{2}, and four \tile{BurningTile}s, as shown in \figref{fig:running-example-lake}.
\end{exampleblock}

The environment corresponds to the domain metamodel that has been previously defined by the domain expert. We see two ways to establish this correspondence. In a native mode, RL experts can define their environments through the API of the chosen RL framework and subsequently, a translation to an instance model typed by the domain metamodel is required. In an MDE mode, RL experts are provided with a suitable DSL to define their environments directly as an instance of the domain metamodel. (See \secref{sec:discussion}.)

\subsection{Configuration of the search process}

In \step{3}, the RL expert configures the search process by (i) defining the mutation operators or mutable states for the automated generation of mutation operators; (ii) defining the constraints the environments and the family must adhere to; (iii) defining the (domain-specific) measures for the objective function that drives the search for the optimal family of environments; and (iv) configuring the GA hyperparameters.
We provide a pluggable architecture with well-defined interfaces to implement when defining components (i)--(iii) of the search process.

\subsubsection{Defining or generating mutations}\label{sec:approach-mt}
Mutations are defined in terms of model transformations (MTs), specified on the domain metamodel. Together with the model of the initial environment, these transformations span the entire search space, i.e., the set of all environment models reachable by applying mutation operators to the initial environment model.

\begin{exampleblock}{Defining mutations}%
The expert defines a model transformation rule \code{changeBurning-} \code{ToRoad}. The LHS matches a \tile{BurningTile} in the \tile{Forest}, and the RHS replaces it with a \tile{RoadTile} at the same coordinates \cell{x}{y}, as shown below.
Applying this rule decreases burning tile density, potentially increases the number of feasible paths in the environment---i.e., decreases the complexity of the learning task.

\begin{lstlisting}[language={VQL}, caption={Example \viatra{} graph query (``Burning tile'')}, label={lst:query-example}]
pattern burningTile(forest : Forest, tile : BurningTile) {
	Forest.states(forest, tile);
}
\end{lstlisting}

\begin{lstlisting}[language={Xtend}, caption={Example mutation expressed as a \viatra{} MT rule that reacts to the match of the graph query in \lstref{lst:query-example}}, label={lst:mutation-example}]
val changeBurningToRoad =
  createRule(burningTile).name("changeBurningToRoad").action[
    val newTile = createRoadTile
    forest.replaceTile(tile, newTile)].build
\end{lstlisting}
\end{exampleblock}

Alternatively, the RL expert may specify the set of mutable states $S_{M}$ to generate the full graph of mutations $S_{M} \to 2^{S_M}$.

\subsubsection{Defining constraints}
\label{sec:approach-constraints}
The RL expert defines the constraints that all generated environments in the family must satisfy. These constraints are used for the validation of environments produced during the environment generation process, filtering out infeasible candidates.
Constraints are expressed as graph queries and subsequently, registered through the ConstraintService. This results in a validation and filtering step in the GA process. Constraints are checked on new mutations. Upon violating a constraint, the mutating model transformation is rolled back through EMF's TransactionalEditingDomain facility---a component in EMF that borrows transactional semantics to reading and writing model resources.

As a requirement, we demand that every environment is solvable.

\begin{exampleblock}{Defining constraints}%
The expert specifies that some states are to be preserved in every mutation:
the \tile{StartTile} is to be kept at \cell{0}{0}, the \tile{GoalTile} at \cell{3}{3}, and the \tile{WaterTile} at \cell{2}{2}.
For this, they first specify a graph query that matches on the state.
\begin{lstlisting}[language={VQL}, caption={\viatra{} graph query to define an invariant}, label={lst:query-example-start}]
pattern startTileFixedLocation(forest: Forest, tile: StartTile) {
  Forest.states(forest, tile);
  Tile.x(tile, x);
  Tile.y(tile, y);
  check((x == 0) && (y == 0));
}
\end{lstlisting}
Then, the query is registered as a constraint:
\begin{lstlisting}[language={Xtend}, caption={Graph query registered as constraint}, label={lst:constraint}]
contraintService.registerConstraint(startTileFixedLocation)
\end{lstlisting}
\end{exampleblock}

\subsubsection{Defining the objective function}\label{sec:approach-of}
The objective function that drives the search is based on population-level measures (i.e., measures over a family of environments) which, in turn, are derived from individual-level measures (i.e., measures over specific environments). These measures enable comparison of populations and individuals, respectively. Comparison is important because it enables ordering of populations and individuals, and, by extension, choosing the best-performing population (i.e., family of environments) and organizing it into a learning process (e.g., a sequence of increasing complexity in CL).

Formally, we demand measure $\mu$ that induces partial ordering but we do not demand countable additivity. Examples of such measures include many useful constructs, such as Shannon entropy, belief functions in Dempster–Shafer theory, and Choquet capacity.
Let $\mathcal{E}$ denote the search space, and let $E \in \mathcal{E}$ denote a particular environment. A feature map $\phi:\mathcal{E} \to \mathbb{R}^n$ extracts measurable structures from environments to produce a feature vector $\phi(E)$.
Let $\Sigma_{\phi}$ be the $\sigma$-algebra over the set of all environments $\mathcal{E}$, representing sets of environments whose features satisfy some property. A complexity measure $\mu: \Sigma_{\phi} \to \mathbb{R}_{\ge 0}$ is a function that assigns non-negative real numbers to measurable sets of environments. This measure satisfies the property of non-negativity, $\mu(\emptyset) = 0$.
The measure induces a partial ordering, $E_1 \preceq E_2 \iff \mu(E_1) \le \mu (E_2)$, that allows for formal comparison of populations in the GA (i.e., families of environments, e.g., by Shannon entropy) and individuals (specific environment models, e.g., by complexity).

For population-level measure, typically, some kind of a diversity measure is chosen. For example, in CL, diversity of environments in terms of an appropriate complexity measure is a good indicator of the eventual successful training performance.
For the individual-level measure, typically, some kind of a complexity measure is chosen. For example, in CL, the sole reason to generate environment variants is the high complexity of the initial environment. Measures are defined by the RL expert, but our framework provides some useful measures, too. These include binned Shannon entropy~\cite{cover1991elements} (which is the default population-level measure), Gini coefficient~\cite{cowell2011measuring}, and statistical variance. Individual-level measures are too domain-specific to define them without a firm grasp on the domain concepts and therefore, our framework does not provide such defaults. With that, defining new measures is straightforward in our framework. The RL expert only needs to implement a simple interface that declares a \code{double evaluate()} method to assign a complexity value to populations or environment models; and subsequently, register it when setting up the GA.

\begin{exampleblock}{Individual (environment) measure of complexity}%
In the running example, the expert defines the complexity $c:\mathcal{E} \rightarrow [0,1)$ based on feasible down-right paths from \tile{StartTile} to \tile{GoalTile}. 
Let $P = \binom{2n-2}{n-1}$ be the total number of down-right paths in an $n \times n$ grid, and let $P_E$ be the number of such paths that avoid all \tile{BurningTile}s in environment $E$.
The complexity is defined as:
\[
c(E) = 1 - \frac{P_E}{P}
\]
This measure provides an admissible heuristic that never overestimates difficulty. 
For this running example, $P = \binom{6}{3} = 20$.
As shown in \figref{fig:fire}, three environments have complexity values $c=0,c=0.5$, and $c=0.7$, corresponding to 0, 2, and 4 burning tiles, respectively.

\begin{figure}[h]
    \centering
    \begin{subfigure}{0.3\linewidth}
        \includegraphics[width=\linewidth]{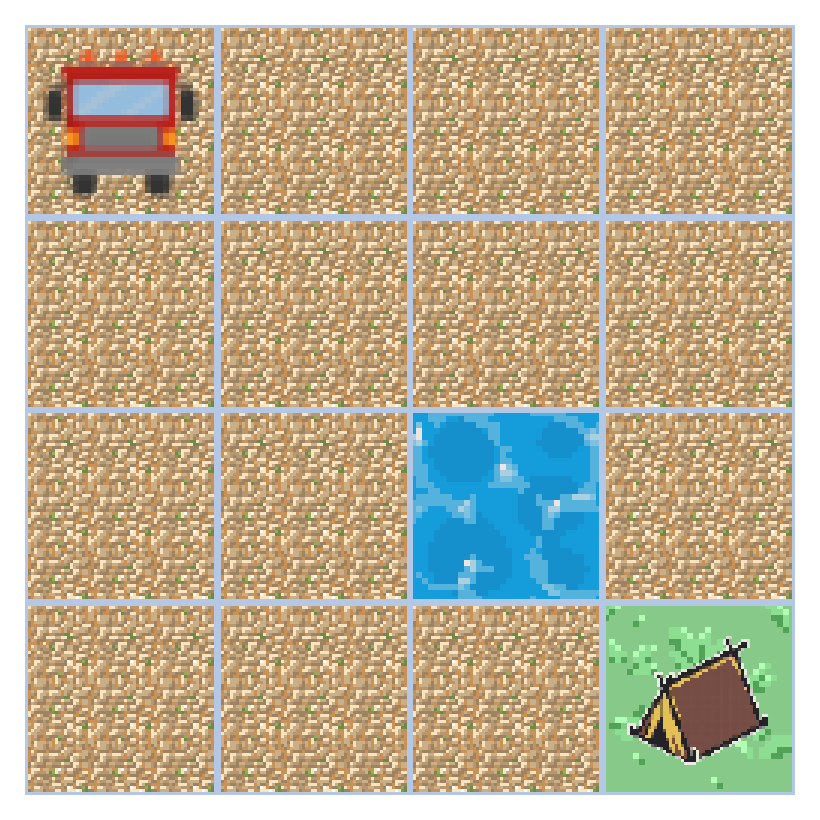}
        \caption{$c = 1 - \frac{20}{20} =0$}
        \label{fig:fire-1}
    \end{subfigure}
    \begin{subfigure}{0.3\linewidth}
        \includegraphics[width=\linewidth]{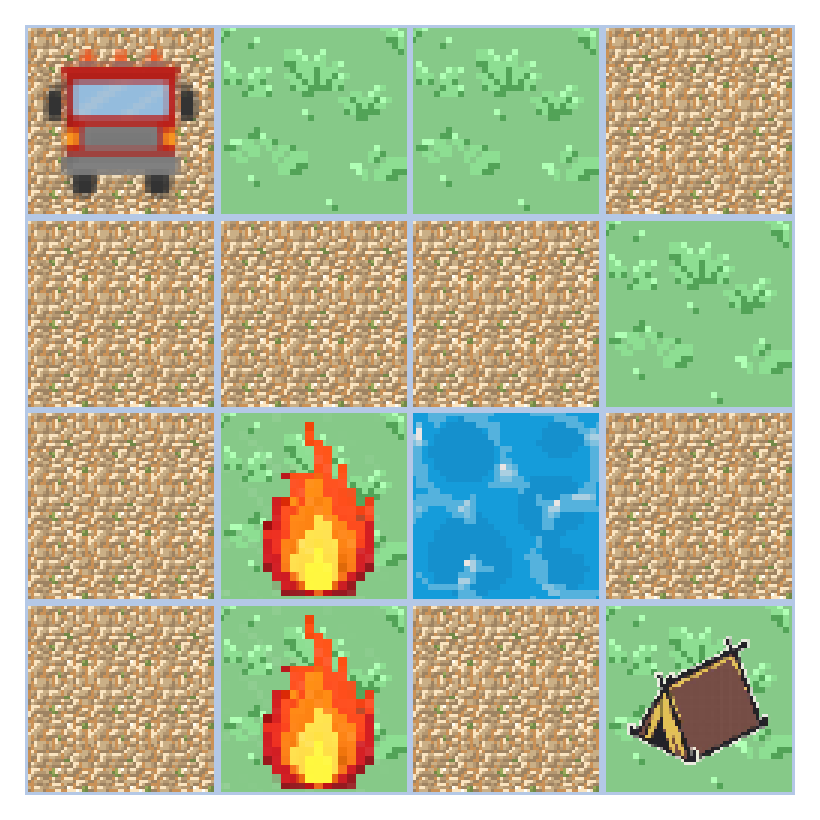}
        \caption{$c = 1 - \frac{10}{20} =0.5$}
        \label{fig:fire-2}
    \end{subfigure}
    \begin{subfigure}{0.3\linewidth}
        \includegraphics[width=\linewidth]{figures/running-example/env_3-complexity_0.70.png}
        \caption{$c = 1 - \frac{2}{20} =0.9$}
        \label{fig:fire-3}
    \end{subfigure}
    \caption{A family of Burning Forest environments in CL}
    \label{fig:fire}
\end{figure}%

\end{exampleblock}

\subsubsection{Defining GA configuration}
The \textit{GA configuration} defines the search process that generates families of environments by applying mutations to the initial environment. This configuration contains algorithm-specific attributes, such as population size, parent and survivor selection mechanisms, mutation rate, maximum mutation attempt threshold, and termination criteria. 

\begin{figure}[h]
    \centering
    \includegraphics[width=0.8\linewidth]{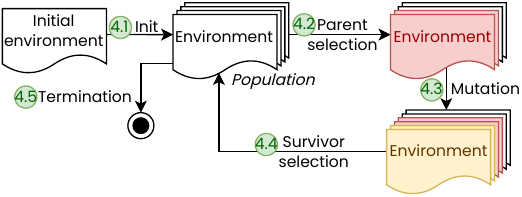}
    \caption{Environment generation process}
    \label{fig:ga-process}
\end{figure}

\subsection{Generating families of environments}
\label{sec:approach-ga}
In \step{4}, a GA generates the family of environments that meet the objective function and respect the defined constraints. (\figref{fig:ga-process})
Since population size directly affects the sampling ability and performance of the GA, limited population size can lead to premature convergence~\cite{tsoy2003influence}. To address this, we use the hybrid GA, a marriage between a population-based global search and a heuristic local search~\cite{moscato1989evolution}, which balances global exploration with local exploitation even with small populations~\cite{el2006hybrid}. In our approach, we use simulated annealing (SA)~\cite{bertsimas1993simulated} as the local search method in the survivor selection phase, maintaining the highest diversity group of individuals for the next generation.
SA is an appropriate choice here as survivor selection, in our case, is a rugged combinatorial problem with a high number of local optima, and SA helps escaping those local optima. This property of SA, in turn, is highly conducive to a better balance between exploration and exploitation. Typical alternative algorithms include hill climbing (for simpler RL problems), beam search (better than hill climbing and the ordered nature of curricula actually supports its optimization semantics), and Monte Carlo Tree Search (for improved long-horizon optimization)~\cite{russell2020artificial}.

\subsubsection{Initialization}
In step \colorgreennumber{4.1} the initial population is created. Clones of the initial environment are created until the specified population size is reached. To introduce diversity, each environment copy is varied using a small, random number of the specified model transformations. After this initial mutation, each environment model is validated using the user-defined constraints, and invalid environments are discarded. 

\subsubsection{Parent selection}
In step \colorgreennumber{4.2}, members of the population are selected to become parents. Each candidate environment model is evaluated using the user-defined objective function. We use random selection but more sophisticated mechanisms can be registered and used in our framework (e.g., tournament selection, elitism).

\subsubsection{Mutation}
In step \colorgreennumber{4.3}, offspring environment models are generated by applying mutations. Each of the parents selected in the previous step is mutated using the mutation operators, with a probability defined by the mutation rate specified in the GA configuration. After mutation, the candidate offspring are validated. If a candidate offspring is invalid or violates previously defined constraints, the model editing transaction rolls back using EMF's TransactionalEditingDomain.
Mutation is reattempted on the parent of the invalid offspring until a valid offspring is produced, or a maximum mutation attempt threshold is met, as specified in the GA configuration. 

\subsubsection{Survivor selection}
In step \colorgreennumber{4.4}, the next generation is formed by performing survivor selection. Each new environment is evaluated for complexity, and a local search algorithm identifies the individuals that together form the population with the highest fitness, e.g., diversity. These individuals are retained and together, they form the new population. We use simulated annealing~\cite{bertsimas1993simulated} as our local search to select survivors, but this can be changed by the user by registering new local search heuristics (e.g., hill climbing, depth-first search, or breadth-first search).

\subsubsection{Termination}
The GA continues until the termination criteria are met (\colorgreennumber{4.5}). These termination criteria are specified in the GA configuration (e.g., maximum number of generations, or timeout).

\subsection{Code generation}
In step \colornumber{5}, the code for the target RL training environment is generated. For this step, the domain expert needs to develop the code generation templates, possibly in collaboration with the RL expert.
\section{Evaluation}\label{sec:evaluation}

To evaluate our approach in a curriculum learning setting on the scaled-up version of the running example. CL is a representative example of the various learning paradigms the rely on families of training environments.
We assess our approach by answering the following research question: \textbf{how do curricula generated in our approach improve learning performance?}

\subsection{Experiment setup}

\subsubsection{Metrics}\label{sec:evaluation-metrics}
We evaluate the RL agent's \textbf{cumulative reward} realized throughout the learning process, as well as the \textbf{success rate} of the trained agent in the target environment. 
Cumulative reward measures the total reward collected over training, which is a standard metric for assessing RL performance~\cite{sutton1998reinforcement}.
Higher cumulative rewards indicate better learning performance. Faster convergence indicates better learning dynamics.
After training, we evaluate each saved policy on the target environment ($E_6$) for $3\,000$ episodes, reporting the mean episodic return (i.e., average cumulative reward), and the average success rate.
The success rate is defined as the proportion of episodes in which the agent reaches the goal while avoiding burning tiles, which is an accepted metric for evaluating task completion in goal-oriented RL~\cite{mclean2025meta-world+}.

\subsubsection{Environment configuration}
We use a scaled-up, $8 \times 8$ version of the running example (\secref{sec:running-example}). This size is sufficient for our purposes as the combinatorial explosion places the problem beyond the humanly tractable horizon and invokes the need for our approach.

In CL, the initial environment is the hardest one, i.e., the target environment (\figref{fig:fire-6}); and the easier ones are to be generated (\figref{fig:fire-1}--\figref{fig:fire-5}).
All MTs and constraints follow those defined in \secref{sec:approach-mt} and \secref{sec:approach-constraints}, respectively.
The complexity measure is defined in relation to the number of available paths between the start and the goal, as defined in \secref{sec:approach-of}.
The reward structure is the same across environments in the family and follows \tabref{tab:rewards-forest}.
As the diversity measure, we use Shannon entropy~\cite{shannon1948mathematical}, a widely adopted measure of diversity in CL~\cite{wang2022survey}, computed over binned complexity values across the population. The complexity is partitioned into $|\mathcal{E}|$ bins, where $|\mathcal{E}|$ is the number of generated environments in the curriculum. Higher entropy indicates a more uniform distribution of complexity values across the family, i.e., higher diversity.

\subsubsection{GA configuration}
We use a population size of six, i.e., the GA will produce a curriculum of six environments. The mutation rate is set to 0.85 to encourage exploration of different environment configurations while preserving some individuals unchanged.
We configure simulated annealing with 500 rounds, an initial temperature of $T_0 = 0.1$, and a cooling rate of $cr = 0.95$ (widely used default, see, e.g., MATLAB\footnote{\url{https://www.mathworks.com/help/gads/simulated-annealing.html}}).
Temperature decreases as $T_{i+1} = \alpha \cdot T_i$. 
The GA terminates after $1\,000$ generations.

\subsubsection{RL configuration}
For RL, we use Q-learning~\cite{watkins1992q-learning}, a fundamental model-free value-based algorithm, which is well-suited for discrete state and action spaces, such as the Burning Forest. 
The learning rate $\alpha$ affects how much the agent updates its policy parameters during training. We use a moderate learning rate ($\alpha = 0.1$) to ensure stable convergence.
The discount factor $\gamma$ determines how much the agent values future rewards. We use a high discount factor ($\gamma = 0.99$), which encourages the agent to prioritize long-term rewards over immediate ones.
For exploration, we use an $\epsilon$-greedy strategy, at each step, the agent selects a random action with probability $\epsilon$ and the greedy action otherwise. $\epsilon$ decays from fully random ($\epsilon = 1$) over training.
This strategy ensures broad exploration in the early stage and gradual exploitation as the agent's Q-values stabilize. 
When the agent advances to a new environment in the curriculum, we reset $\epsilon$ to $1$, encouraging the agent to re-explore the new environment. These hyperparameters are common defaults in RL practice; we use them to mitigate threats to the internal validity.

\subsubsection{CL configuration}
From the six generated environments ($E_1$ $\rightarrow$ $E_2$, ... $\rightarrow$ $E_6$) ordered by complexity, we construct five curricula with different prefixes: $\{E_1 \rightarrow ... \rightarrow E6\}$, $\{E_2 \rightarrow ... \rightarrow E6\}$, $\{E_3 \rightarrow ...\rightarrow E6\}$, $\{E_4 \rightarrow E_5 \rightarrow E6\}$, and $\{E_5 \rightarrow E6\}$.
Each environment is trained for $50\,000$ steps. The total training budget scales with the length of the curriculum. 
Baseline agents train directly on each environment for $300\,000$ steps, the budget of the full curriculum.

\subsection{Results and key observations}

\begin{figure}[h]
    \centering
    \begin{subfigure}{0.3\linewidth}
        \includegraphics[width=\linewidth]{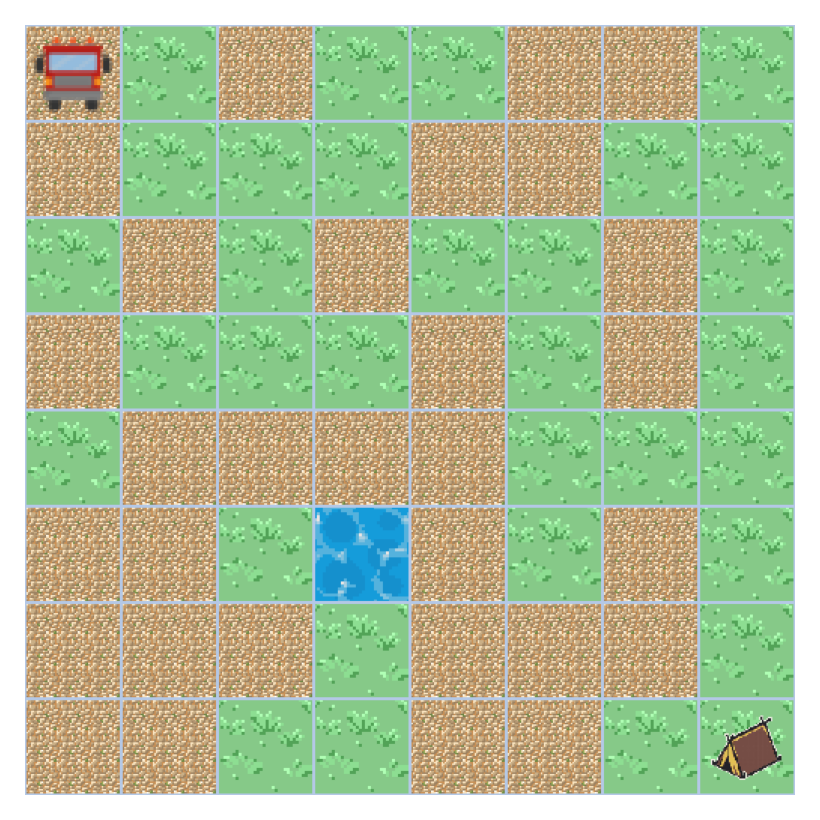}
        \caption{c = 0.00}
        \label{fig:fire-1}
    \end{subfigure}
    \begin{subfigure}{0.3\linewidth}
        \includegraphics[width=\linewidth]{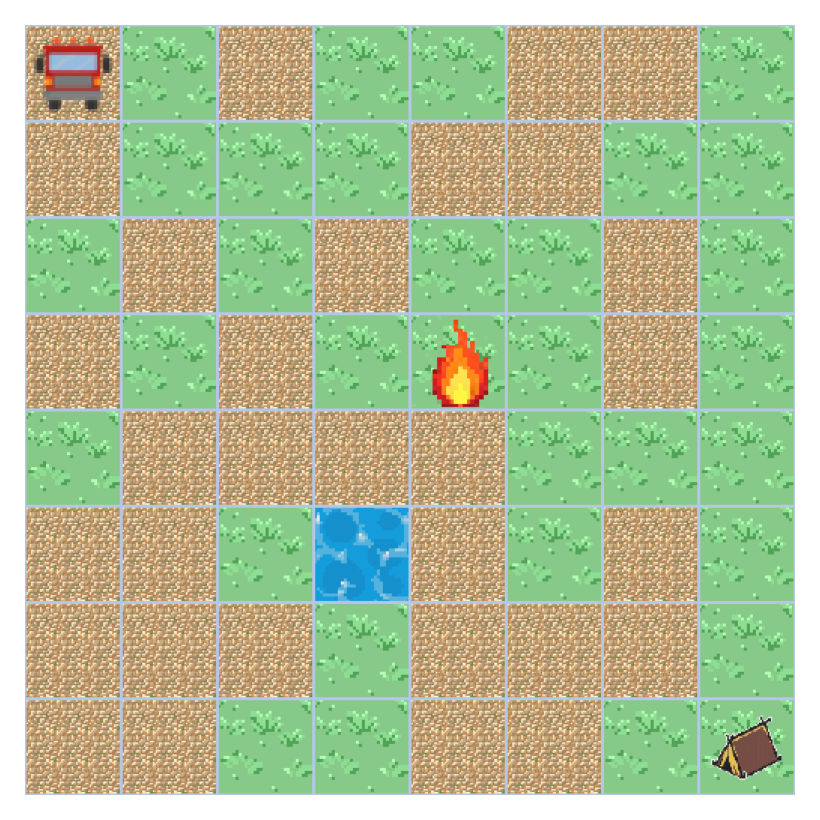}
        \caption{c = 0.20}
        \label{fig:fire-2}
    \end{subfigure}
    \begin{subfigure}{0.3\linewidth}
        \includegraphics[width=\linewidth]{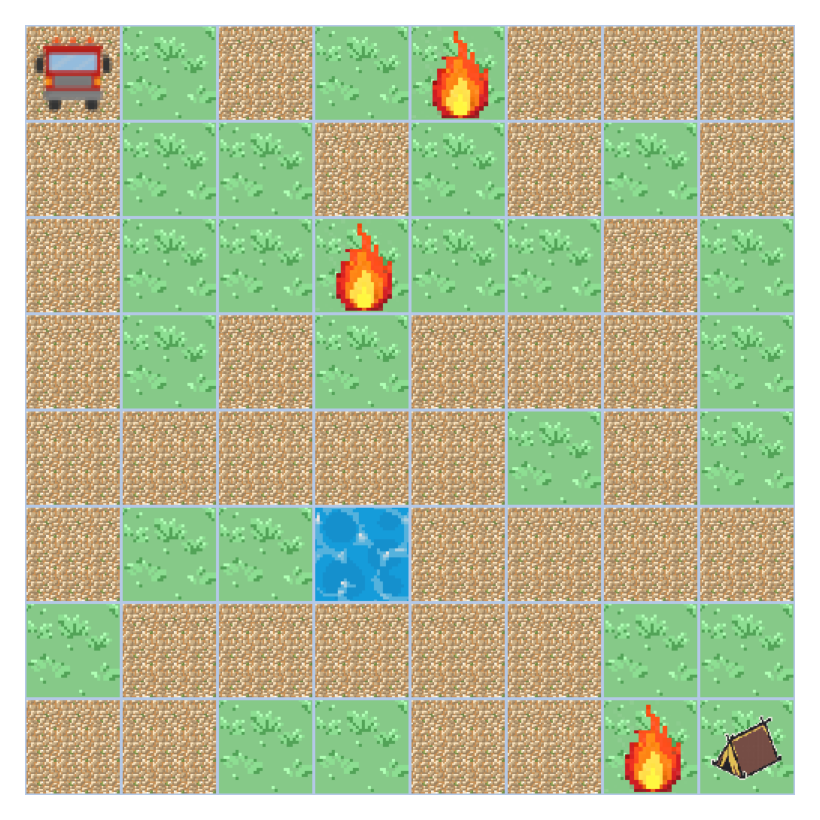}
        \caption{c = 0.48}
        \label{fig:fire-3}
    \end{subfigure}
    \begin{subfigure}{0.3\linewidth}
        \includegraphics[width=\linewidth]{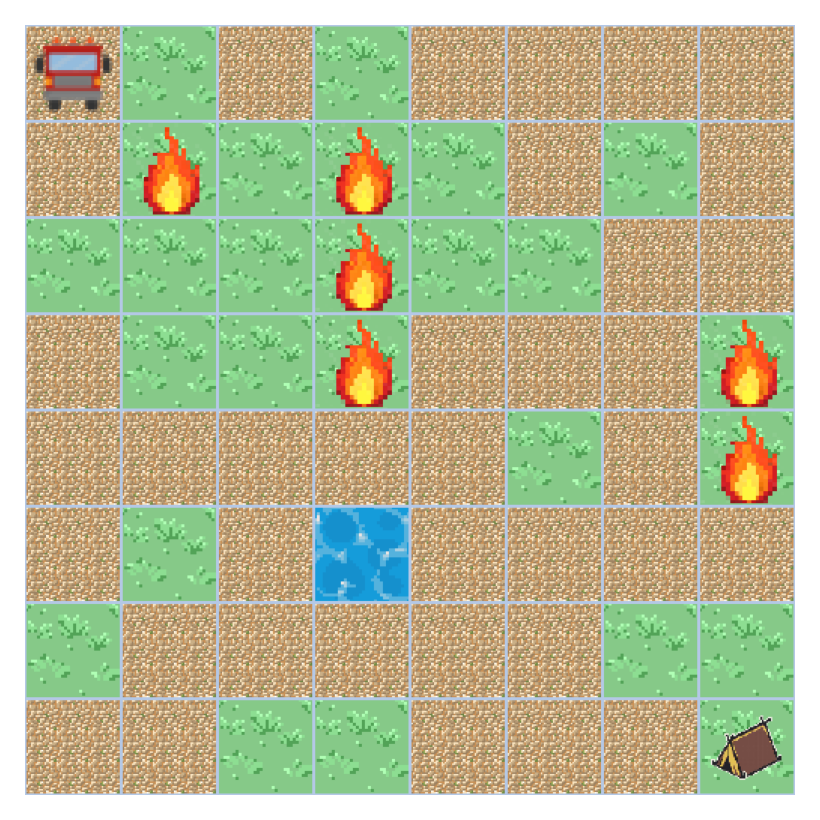}
        \caption{c = 0.59}
        \label{fig:fire-4}
    \end{subfigure}
    \begin{subfigure}{0.3\linewidth}
        \includegraphics[width=\linewidth]{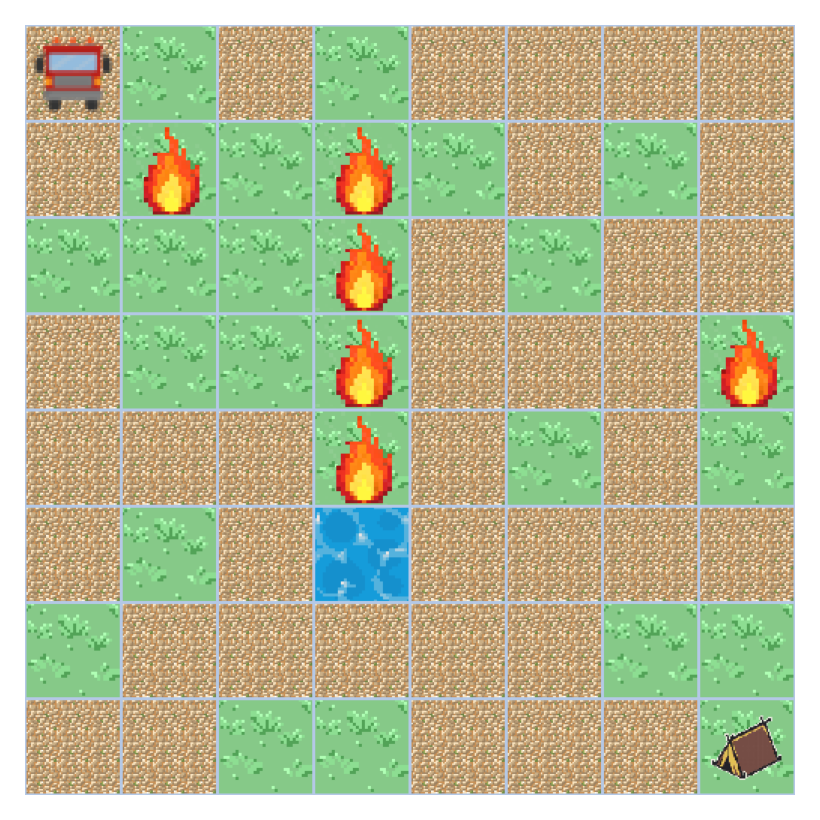}
        \caption{c = 0.67}
        \label{fig:fire-5}
    \end{subfigure}
    \begin{subfigure}{0.3\linewidth}
        \includegraphics[width=\linewidth]{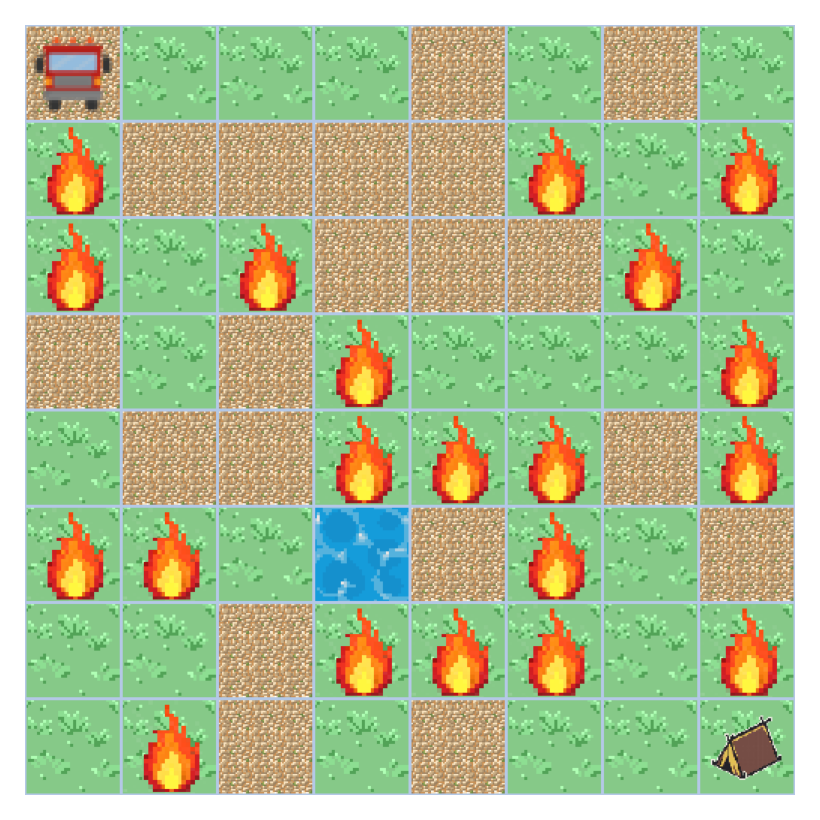}
        \caption{c = 0.93}
        \label{fig:fire-6}
    \end{subfigure}
    \caption{Generated curriculum of increasing complexity}
    \label{fig:generated-environments}
\end{figure}

\figref{fig:generated-environments} visualizes the generated curriculum. Each environment indeed satisfies the defined constraints: a valid path from start to goal exists in each environment, burning tiles are placed within the density range, and complexity increases progressively.

\begin{figure*}[t]
    \centering
    \begin{subfigure}{0.3\textwidth}
        \includegraphics[trim={0 0.25cm 0 0},clip,width=\linewidth]{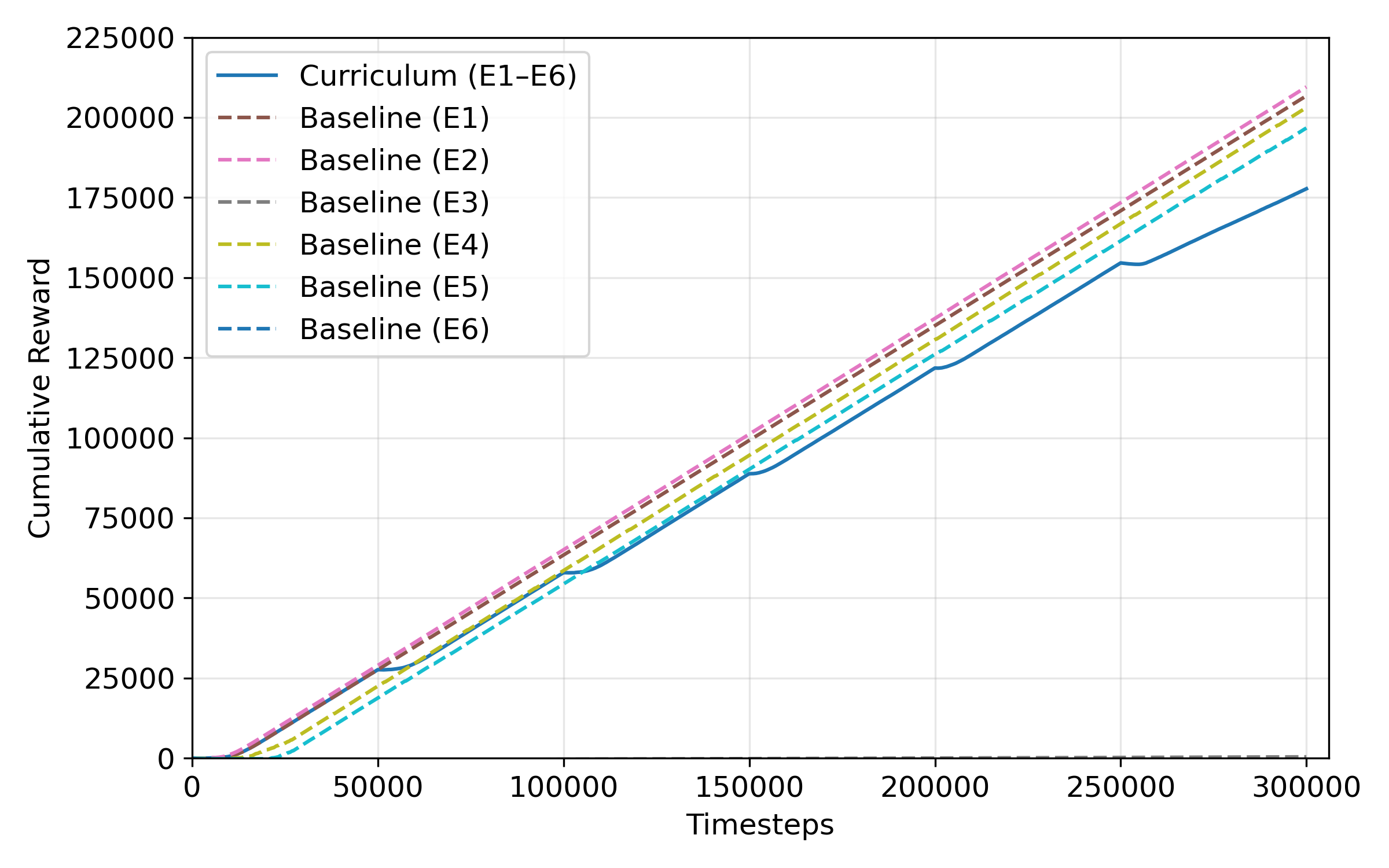}
        \caption{Curriculum prefix E1-E6}
        \label{fig:curriculum-1}
    \end{subfigure}
    \hfill
    \begin{subfigure}{0.3\textwidth}
        \includegraphics[trim={0 0.25cm 0 0},clip,width=\linewidth]{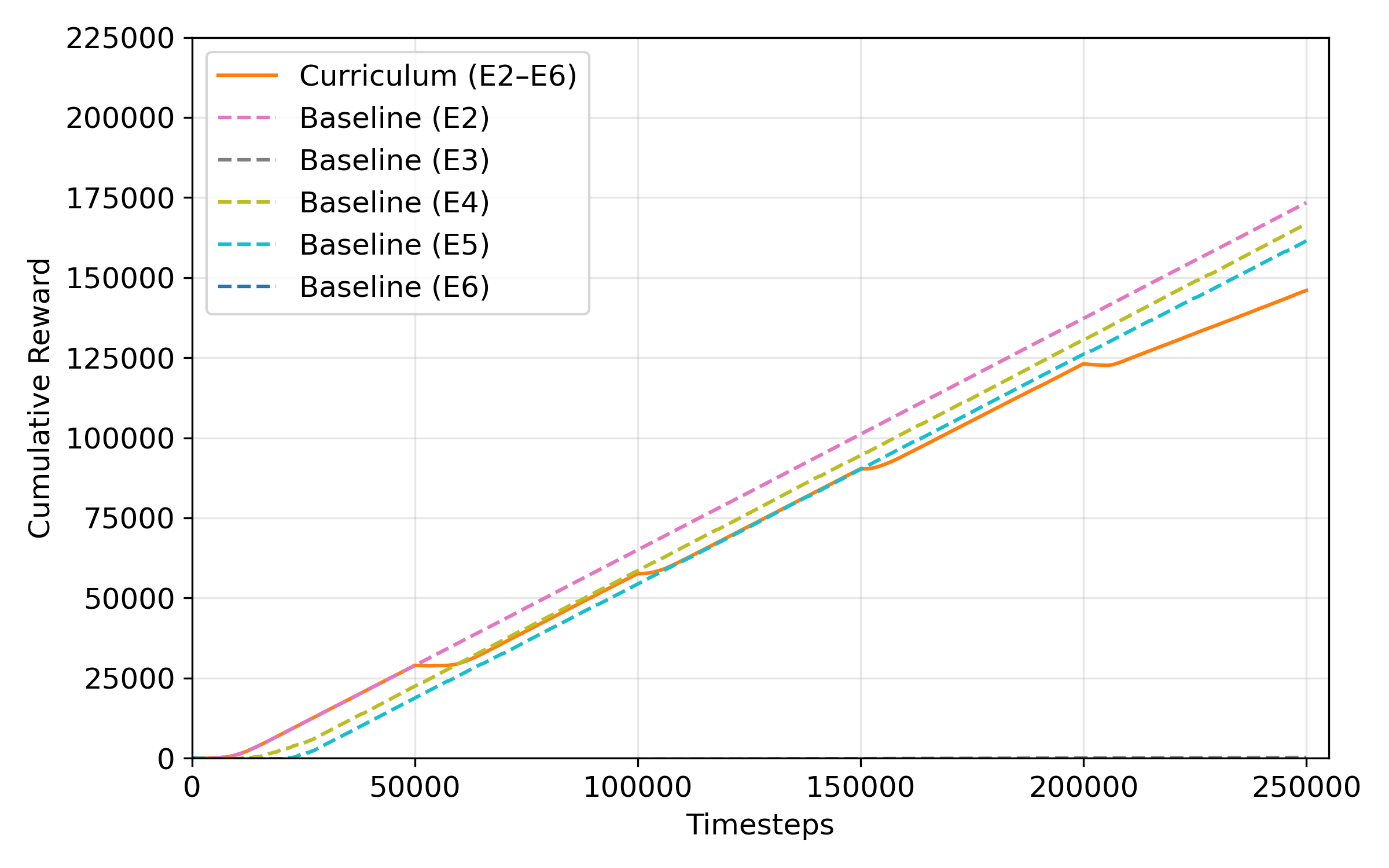}
        \caption{Curriculum prefix E2-E6}
        \label{fig:curriculum-2}
    \end{subfigure}
    \hfill
    \begin{subfigure}{0.3\textwidth}
        \includegraphics[trim={0 0.25cm 0 0},clip,width=\linewidth]{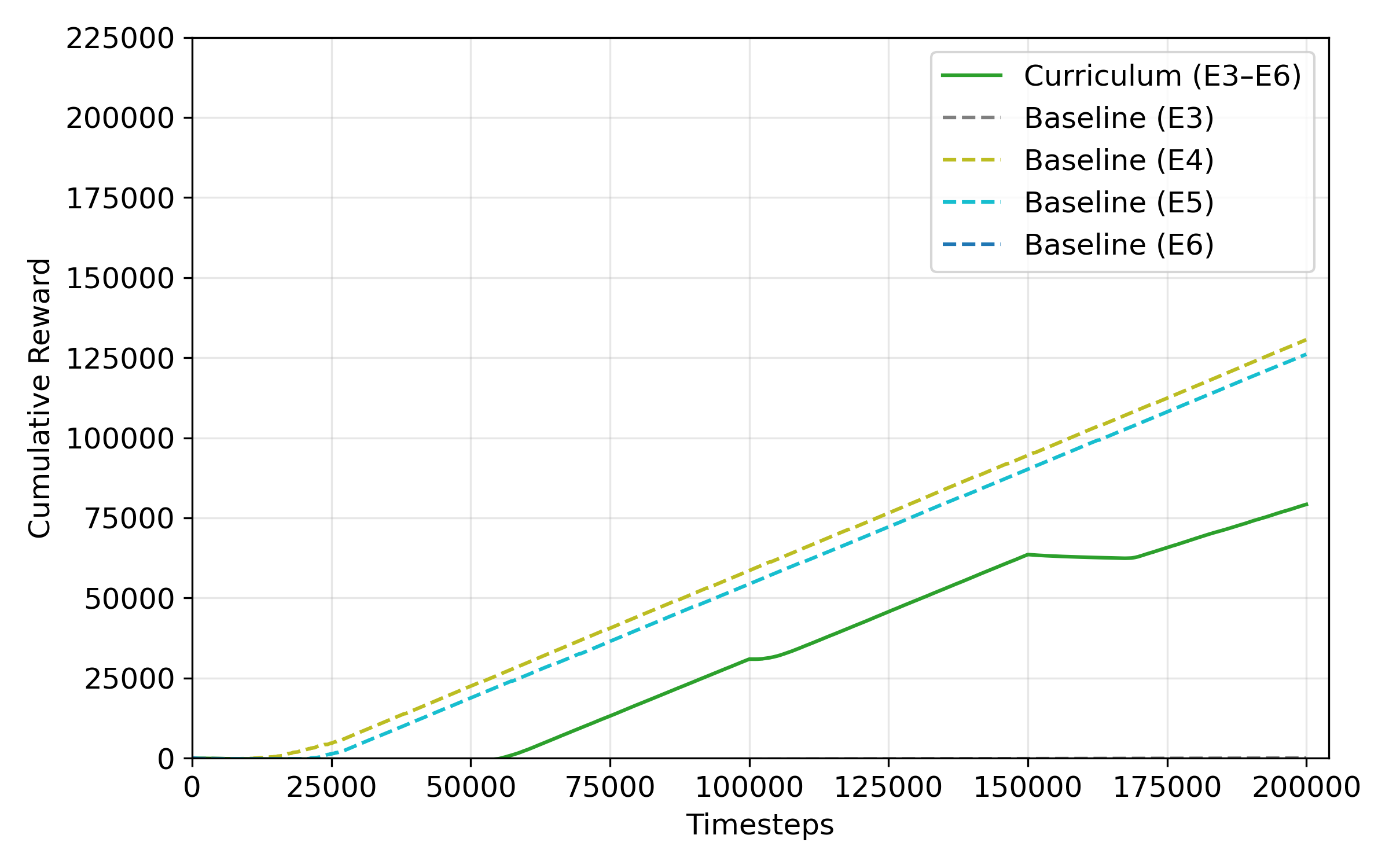}
        \caption{Curriculum prefix E3-E6}
        \label{fig:curriculum-3}
    \end{subfigure}\\[0.5em]
    \begin{subfigure}{0.3\textwidth}
        \includegraphics[trim={0 0.25cm 0 0},clip,width=\linewidth]{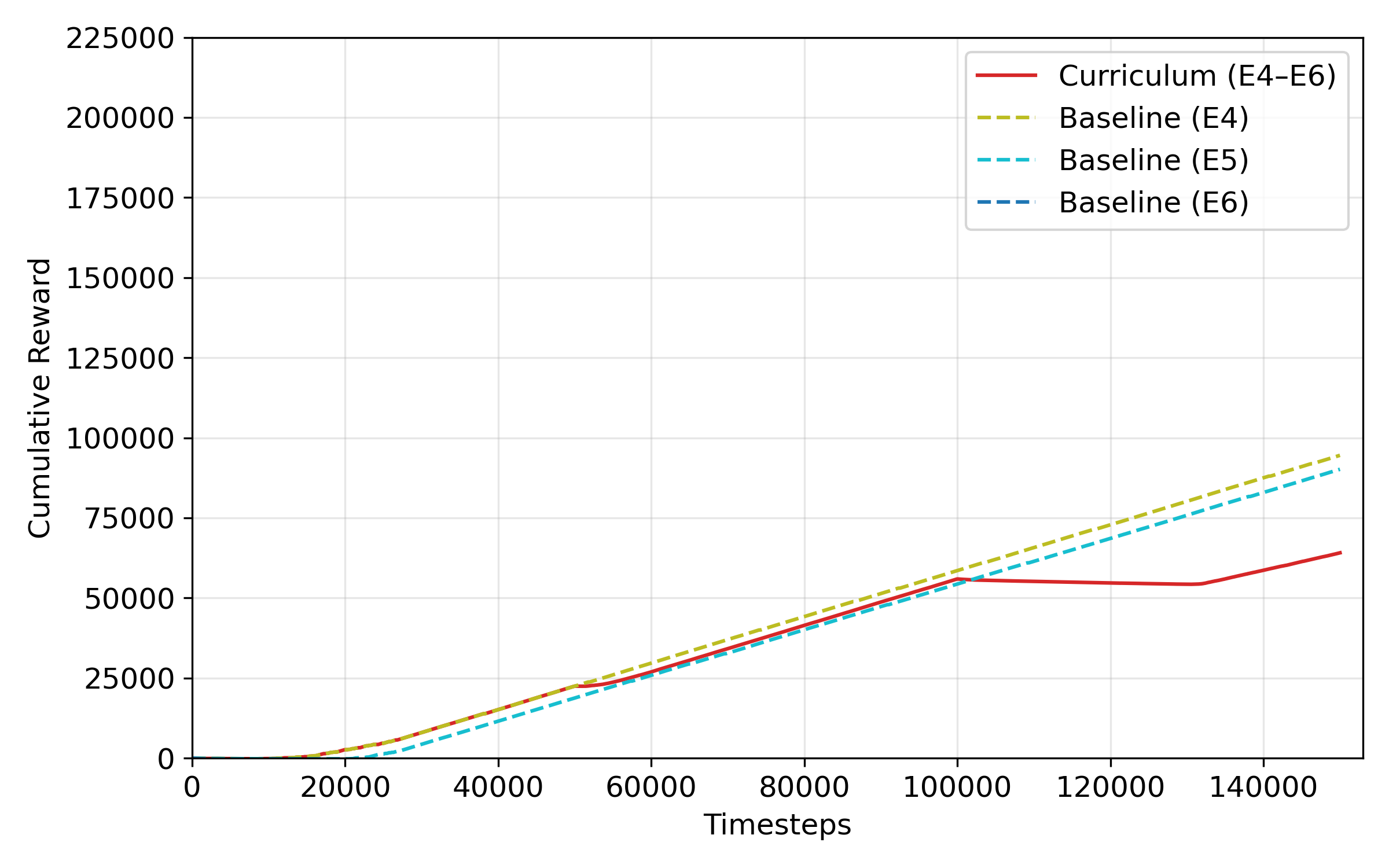}
        \caption{Curriculum prefix E4-E6}
        \label{fig:curriculum-4}
    \end{subfigure}
    \hfill
    \begin{subfigure}{0.3\textwidth}
        \includegraphics[trim={0 0.25cm 0 0},clip,width=\linewidth]{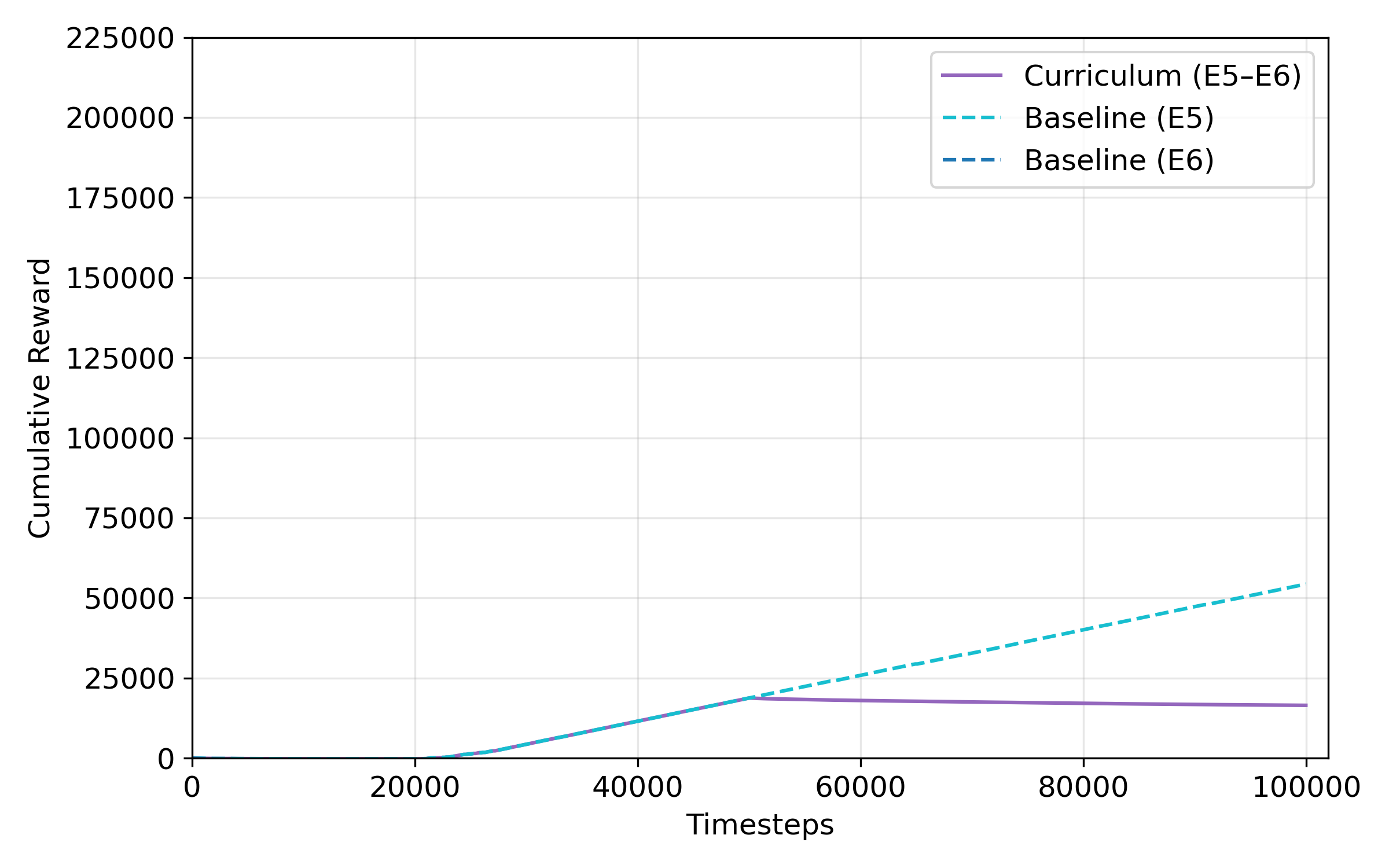}
        \caption{Curriculum prefix E5-E6}
        \label{fig:curriculum-5}
    \end{subfigure}
    \hfill
    \begin{subfigure}{0.3\textwidth}
         \includegraphics[trim={0 0.25cm 0 0},clip,width=\linewidth]{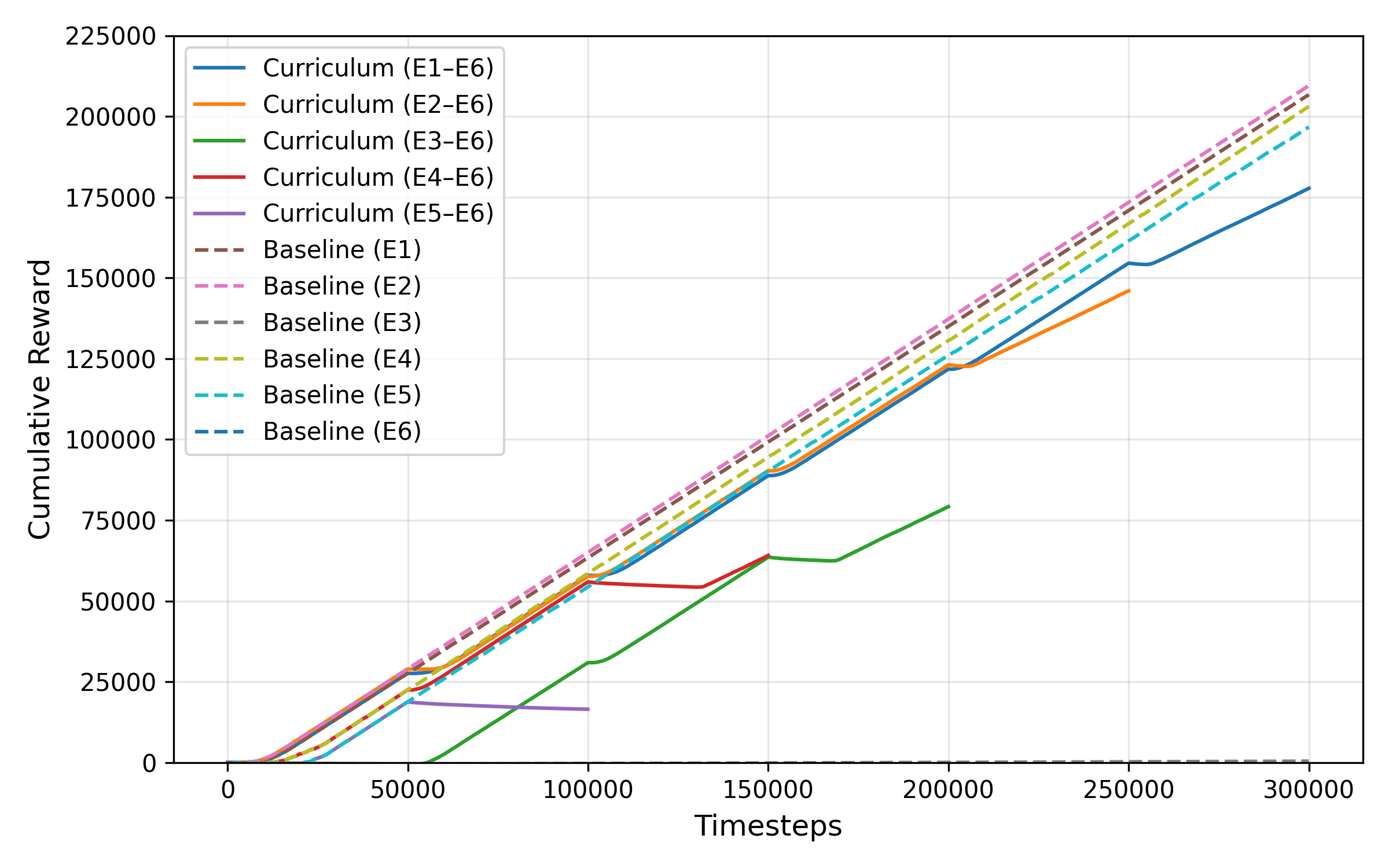}
         \caption{All curricula}
         \label{fig:curriculum-combined}
     \end{subfigure}
    \caption{Cumulative reward with different prefixes. Prefixes shown individually (a--e) and combined (f).}
    \label{fig:curriculum-all}
\end{figure*}

\figref{fig:curriculum-all} shows cumulative rewards during training. 
Each subfigure in \figref{fig:curriculum-all} compares a curriculum agent against baselines trained on the individual environments in that curriculum.
\tabref{tab:evaluation} presents the trained agents' performance in the target environment.

We observe that direct training in a single environment is insufficient without prior exposure to simpler environments, as the agent fails to learn the effective policy.
The agent trained directly on $E_6$ accumulates negative reward throughout training, and achieves $0\%$ success rate and a mean episodic return of $0$ during evaluation.
Although baseline agents trained on environments other than $E_3$ achieve positive rewards and accumulate higher rewards than curriculum-based agents, these learned policies fail to directly transfer to the most challenging environment.
They achieve non-positive mean episodic return and $0\%$ success rate during evaluation.

In contrast, we observe that \textbf{the generated curriculum improves learning performance.} 
All curriculum-trained agents accumulate positive reward during training, and all achieve a $100\%$ success rate on $E_6$, except the agent trained with a curriculum comprising only the two hardest environments (\figref{fig:curriculum-5}).

We observe that curriculum-trained agents exhibit plateaus when transitioning to the new environment, during which the agent adapts its learned policy to the harder configuration.
These plateaus are shorter when the agent has been exposed to more intermediate environments, 
For example, the full curriculum-trained agent (\figref{fig:curriculum-1}) transitions smoothly between environments with short plateaus.
In contrast, shorter curricula,  such as \figref{fig:curriculum-3} and \figref{fig:curriculum-4}, exhibit long plateaus during transitions.
This suggests that finer-grained progression through intermediate environments facilitates faster transfer, as larger gaps between consecutive environments lead to longer adaptation periods.
This aligns with findings in sim-to-real transfer, where greater investment in training on diverse foundational skills produces more robust policies that generalize better to real-world settings~\cite{chaffre2020sim-to-real}.

\begin{table}[t]
    \renewcommand{\arraystretch}{0.7}
    \centering
    \small
    \caption{Evaluation results on the target environment}
    \label{tab:evaluation}
    \small
    \begin{tabular}{@{}lcc@{}}
        \toprule
        \textbf{Agent} & \textbf{Success rate (\%)} & \textbf{Mean episodic return} \\
        \midrule
        Baseline ($E_1$) & 0 & -1.0 \\
        Baseline ($E_2$) & 0 & -1.0 \\
        Baseline ($E_3$) & 0 & -1.0 \\
        Baseline ($E_4$) & 0 & -1.0 \\
        Baseline ($E_5$) & 0 & -1.0 \\
        Baseline ($E_6$) & 0 & 0 \\
        \midrule
        Curriculum ($E_1$--$E_6$) & 100 & 10.2 \\
        Curriculum ($E_2$--$E_6$) & 100 & 10.2 \\
        Curriculum ($E_3$--$E_6$) & 100 & 10.2 \\
        Curriculum ($E_4$--$E_6$) & 100 & 10.2 \\
        Curriculum ($E_5$--$E_6$) & 0 & 0 \\
        \bottomrule
    \end{tabular}
\end{table}

To validate the scalability of our approach, we run the curriculum generation process for different sized problems and curriculum sizes. As shown in \tabref{tab:generation-time}, generation time scales well compared to the combinatorial explosion observed in the state space that is induced by the growing dimensions of the problem (map).
Generation time increases with the curriculum size more substantially. However, this kind of scalability is not in the scope of our work, as we focused on demonstrating the feasibility of using MDE for generating environment families. The scalability in terms of curriculum size can be addressed in multiple ways in subsequent works, e.g., by incrementalization of the end-to-end GA process, or by choosing better-scaling local search and complexity assessment algorithms.

\begin{table}[t]
    \centering
    \setlength{\tabcolsep}{3pt}
    \renewcommand{\arraystretch}{0.8}
    \caption{Generation time (seconds) for different problem sizes}
    \label{tab:generation-time}
    \small
    \begin{tabular}{@{}cccccc@{}}
    \toprule
        Map size & 8$\times$8 & 9$\times$9 & 10$\times$10 & 11$\times$11 & 12$\times$12 \\
        State space & \num{3.4e30} & \num{4.4e38} & \num{5.2e47} & \num{5.4e57} & \num{5.0e68} \\
        \hline
        Curriculum size & \multicolumn{5}{c}{Generation time} \\
        6  & 47  & 49 & 49 & 49 & 50 \\
        10 & 208 & 211 & 212 & 215 & 225 \\
        20 & 1594  & 1718 & 1617 & 1642 & 1659 \\
    \bottomrule
    \end{tabular}
\end{table}

\subsection{Evaluation against random baselines}

To illustrate that the improvement comes from the curriculum's structure and not merely from exposure to multiple environments, we compare our generated curriculum against two randomized baselines: randomly generated environment sequences and the same sequences reordered by our complexity measure, together with direct training on the target environment ($E_6$).
Each baseline sequence has six environments, matching the count used by our generated curriculum, so the comparison isolates ordering from environment count.
To capture how much randomized sequences vary, we generate 30 baseline sequences and train an agent on each. The randomized baseline curves report the mean over these 30 runs, with shaded regions indicating $\pm 1$ standard deviation. We repeat the comparison under three training budgets, $10\,000$, $50\,000$, and $100\,000$ steps per environment, to check whether the result holds independent of training length.

\begin{figure*}[t]
    \centering
    \begin{subfigure}{0.3\textwidth}
        \includegraphics[trim={0 0.25cm 0 0},clip,width=\linewidth]{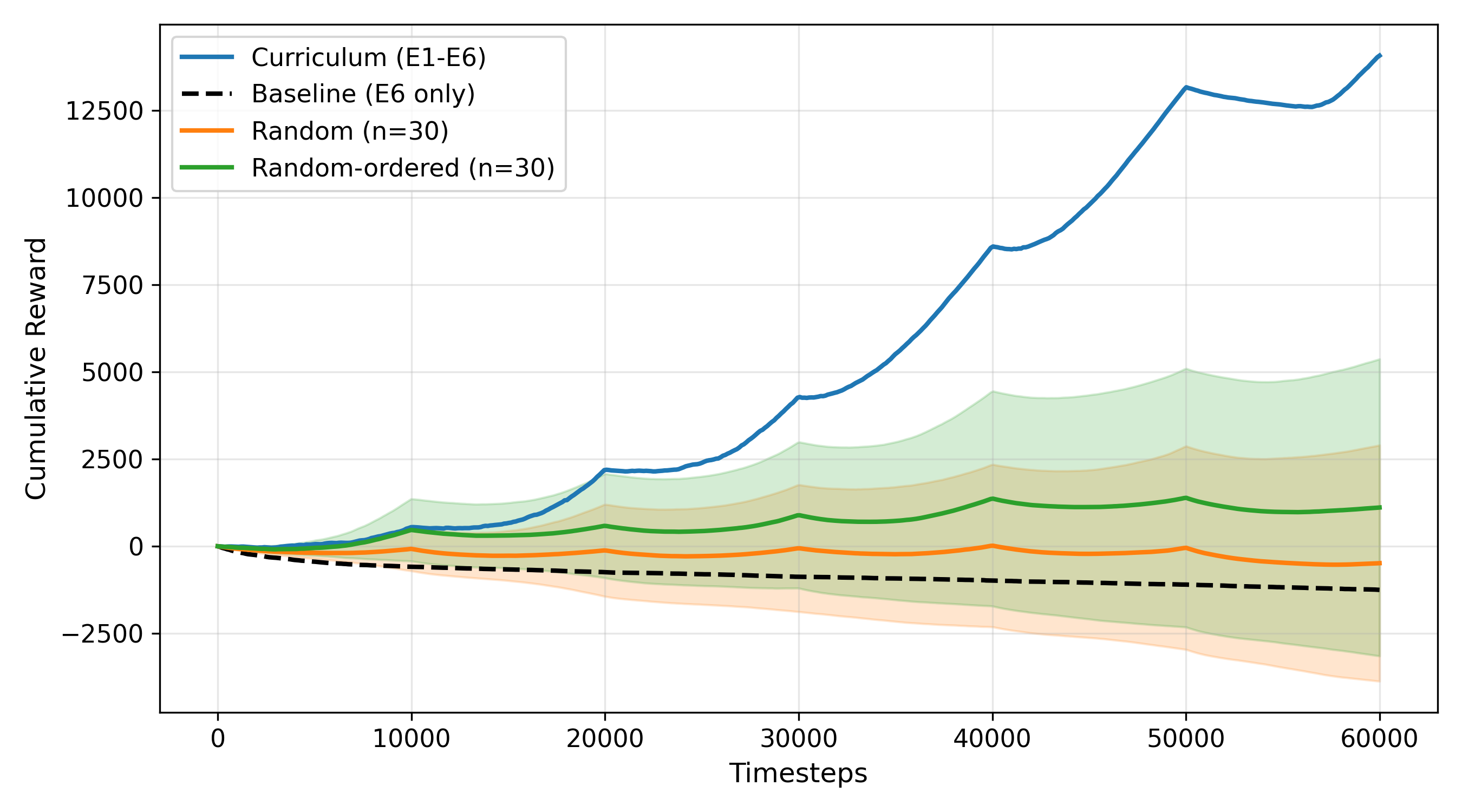}
        \caption{Shorter (10k steps/env)}
        \label{fig:random-baselines-shorter}
    \end{subfigure}
    \hfill
    \begin{subfigure}{0.3\textwidth}
        \includegraphics[trim={0 0.25cm 0 0},clip,width=\linewidth]{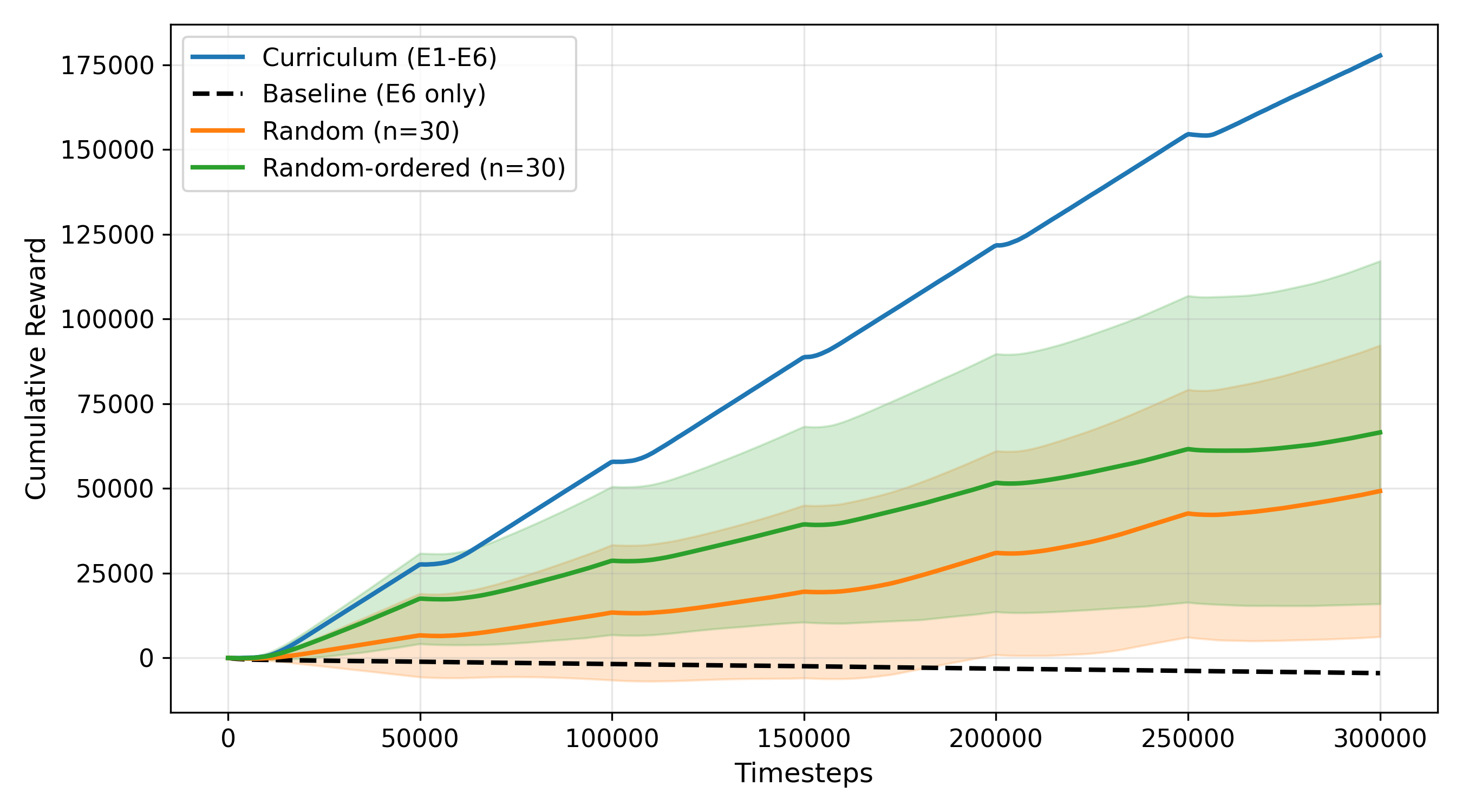}
        \caption{Normal (50k steps/env)}
        \label{fig:random-baselines-normal}
    \end{subfigure}
    \hfill
    \begin{subfigure}{0.3\textwidth}
        \includegraphics[trim={0 0.25cm 0 0},clip,width=\linewidth]{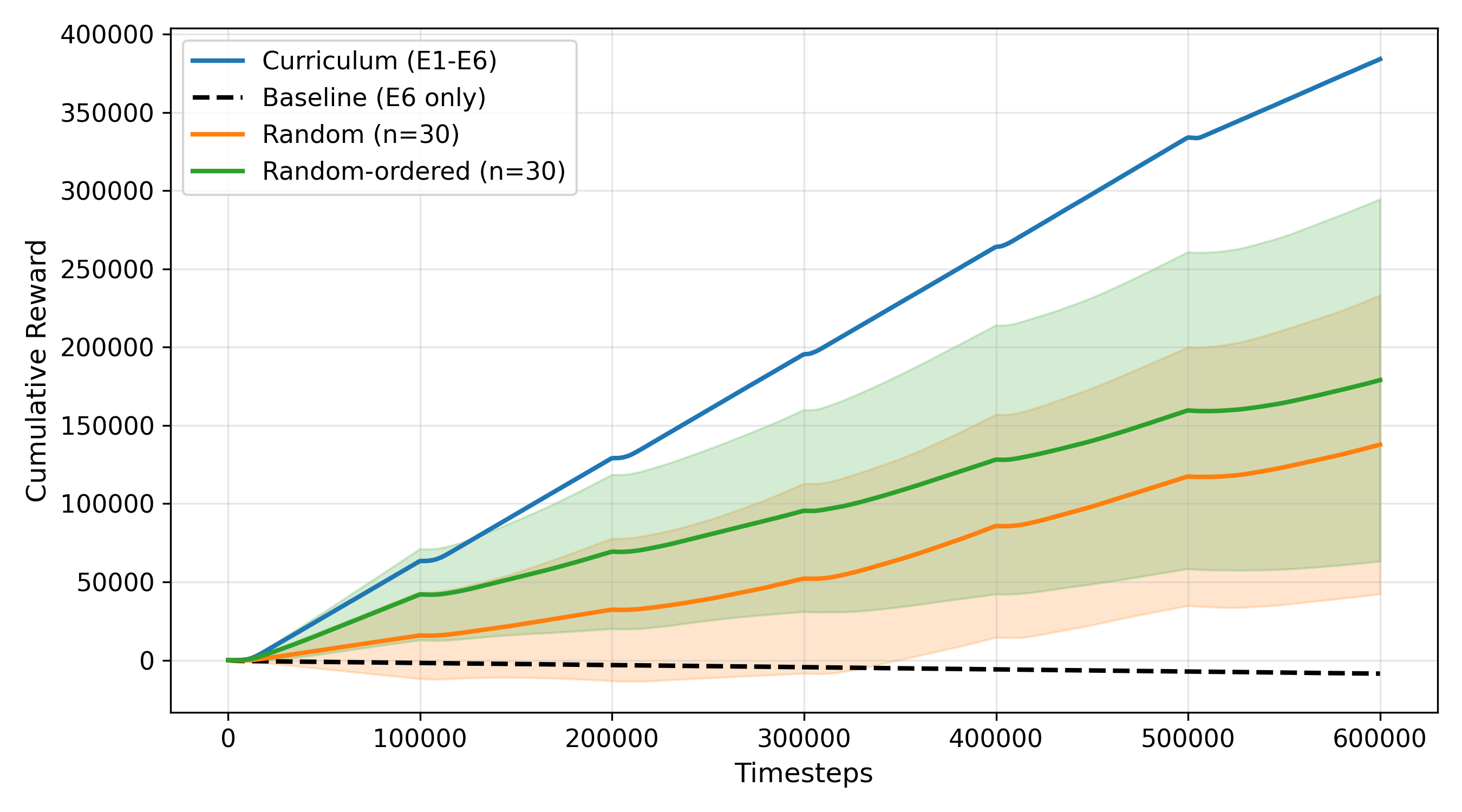}
        \caption{Longer (100k steps/env)}
        \label{fig:random-baselines-longer}
    \end{subfigure}\\[0.5em]
    \caption{Cumulative reward for generated curriculum against random baselines, across different training budgets}
    \label{fig:random-baselines}
\end{figure*}

\figref{fig:random-baselines} shows cumulative reward during training for all four settings at each budget. We observe that \textbf{the generated curriculum accumulates reward faster and more consistently than random baselines at every budget.}
\tabref{tab:random-baseline-evaluation} reports mean episodic return and success rate on $E_6$ after training. We observe that \textbf{the generated curriculum reaches a 100\% success rate across all budgets}, and the random baselines never match it. 
Under the shorter budget, random sequences succeed only $10\%$ of the time and ordered random sequences $13\%$. 
Even under the longer budget, the average success rate the random baselines reach is $63\%$, far below the $100\%$ achieved by the generated curriculum.
Ordering the random sequences by complexity increases the average success rate slightly, but the variance stays as high as the unordered baseline.

Overall, our results show that \textbf{across all tested training budgets, the proposed curriculum produces more stable and better learning than the randomized baselines}.
The results also indicate that the structure of the generated curriculum matters, beyond simply exposing the agent to multiple random environments.

\begin{table}[t]
    \renewcommand{\arraystretch}{0.7}
    \centering
    \small
    \caption{Target environment against random baselines}
    \label{tab:random-baseline-evaluation}
    \small
    \begin{tabular}{@{}lcc@{}}
        \toprule
        \textbf{Agent} & \textbf{Success rate (\%)} & \textbf{Mean episodic return} \\
        \midrule
        \multicolumn{3}{l}{\textit{Shorter (10k steps/env)}} \\
        Direct training ($E_6$) & 0 & 0 \\
        Random sequences & 10 & 0.67 \\
        Ordered random sequences & 13 & 0.60 \\
        Curriculum & \textbf{100} & \textbf{10.2} \\
        \midrule
        \multicolumn{3}{l}{\textit{Normal (50k steps/env)}} \\
        Direct training ($E_6$) & 0 & 0 \\
        Ordered random sequences & 43 & 3.91 \\
        Random sequences & 47 & 4.37 \\
        Curriculum & \textbf{100} & \textbf{10.2} \\
        \midrule
        \multicolumn{3}{l}{\textit{Longer (100k steps/env)}} \\
        Direct training ($E_6$) & 0 & 0 \\
        Random sequences & 63 & 5.98 \\
        Ordered random sequences & 63 & 6.30 \\
        Curriculum & \textbf{100} & \textbf{10.2} \\
        \bottomrule
    \end{tabular}
\end{table}

\subsection{Threats to validity}

\textit{Internal validity.}
The choice of hyperparameters may influence the observed results. We use established settings for Q-learning in our experiments to mitigate threats to the internal validity.
The reward structure may influence agent performance. To mitigate this, we use a dense reward structure to benefit both CL and direct training by providing shaping signals.

\textit{Construct validity.}
We evaluate the RL agent's cumulative reward realized throughout the learning process. However, cumulative reward may not reflect policy quality, i.e., an agent may converge to a suboptimal policy. To mitigate this, we also evaluate the success rate and average episodic return in the target environment and report statistics.
The choice of complexity measure influences the ordering of environments in the curriculum, which may affect learning performance. To mitigate this, we use an admissible heuristic that never overestimates difficulty -- the number of valid down-right paths from start to goal, where fewer feasible paths indicate higher complexity.
The choice of diversity measure influences the population of generated environments. To mitigate this, we use Shannon entropy~\cite{shannon1948mathematical}, a widely used measure of diversity.

\textit{External validity.}
Our evaluation focuses on CL as the learning paradigm. However, our framework is designed in a way that it supports any learning paradigm as long as population- and individual-level measures are properly defined. Thus, we are reasonably confident that our approach generalizes safely to other paradigms, such as multi-task learning~\cite{sodhani2021multi-task}, meta-RL~\cite{hospedales2022meta-learning}, and domain randomization~\cite{tobin2017domain}. We recommend replication studies to test this claim.
\section{Discussion}\label{sec:discussion}

In this section, we reflect on the approach and set forth important research directions for the MDE community.

\subsection{Reflections and takeaways}

\subsubsection{Curriculum size matters}
Our evaluation results show that curriculum effectiveness does not scale monotonically with the number of environments. 
The agent trained through five environments (\figref{fig:curriculum-2}) achieves $100\%$ success rates, converging quickly with short plateaus between environments. Agents trained in three (\figref{fig:curriculum-4}) and four (\figref{fig:curriculum-3}) environments also reach the goal successfully, although they exhibit longer plateaus during transitions to harder environments.
This suggests a saturation point, i.e., beyond a sufficient number of intermediate environments, additional environments produce diminishing returns~\cite{shephard1974law}.
This observation implies that both the quality and the quantity of generated environments matter, also noted by \citet{narvekar2020curriculum}. 
Rather than maximizing the number of generated environments, engineers benefit from methods that identify a compact, effective curriculum. 
This motivates three directions: (i) local search strategies that iteratively remove redundant environments to downsize the generated pool; (ii) approximation methods that can estimate curriculum quality without full simulation to enable early pruning; and (iii) incremental environment generation methods that allow engineers to iteratively refine curricula within realistic time and resource constraints.

\subsubsection{Complexity measures are indeed challenging to define}
Our results demonstrate that a human-crafted complexity measure may not correspond exactly to the learning difficulty of the RL agent. $E_3$ (\figref{fig:fire-3}) with a complexity of 0.48 and consists of only three burning tiles appears moderately difficult. However, as shown in~\figref{fig:curriculum-3}, the agent fails to accumulate positive reward when trained directly on it, and the curriculum-based agent starting from $E_3$ exhibits a long plateau before progressing to the next environment.
This confirms a known challenge of determining environment difficulty for agents in curriculum learning~\cite{narvekar2016source}.
This highlights the need for domain-specific languages that enable engineers to express and compose complexity measures from domain primitives. Developing such languages, potentially generated from domain ontologies~\cite{elaasar2023opencaesar}, may be an important future direction.

\subsubsection{Applicability in other learning paradigms}

While our evaluation uses CL as an example, our framework can be applied in other learning paradigms, too.

Domain randomization (DR)~\cite{tobin2017domain} is a widely adopted technique to bridge the sim-to-real gap. Instead of relying on the high-fidelity of simulation during training, DR exposes agents to varied conditions during training, thereby promoting generalization and enhancing the agent's ability to operate reliably in real-world conditions~\cite{pitkevich2024survey}. Our framework can be used to produce environment variants for training~\cite{zhao2020sim-to-real} at a given difficulty range. Our approach also naturally accommodates the combined paradigm of CL and DR, in which agents are simultaneously exposed to both increasing difficulty and increasing randomization~\cite{tobin2017domain}.

Two other paradigms that could benefit from our approach are multi-task RL (MTRL) and meta-RL. In MTRL, a single policy is trained across multiple environments in parallel to maximize average performance~\cite{hospedales2022meta-learning}. In meta-RL, agents learn from a training set of environments and must rapidly adapt to unseen test environments~\cite{hospedales2022meta-learning}.
Both paradigms require environments with consistent observation and action spaces, shared reward structure, and sufficient diversity between environments to enable knowledge transfer rather than memorization.
Our framework supports generating such families by specifying structural invariants while applying mutations to vary task-specific properties (e.g., objects, goals, interaction types), with the diversity measure to ensure adequate task variation for generalization.

Our framework can be of high utility in broader use cases throughout the RL development lifecycle. For testing and robustness evaluation, our framework can be used to generate edge cases and challenging boundary conditions that rarely arise during training but are critical for deployment and safety concerns. For benchmarking, our framework can produce environments with controlled variation in difficulty and diversity, enabling fair and reproducible comparison between RL algorithms. Additionally, the framework can generate a larger population from a small initial population as a reference distribution, while preserving statistical properties (e.g., mean, variance) and using out-of-distribution detection~\cite{liu2023towards} to ensure distribution consistency.

These directions demonstrate that our model-driven approach provides a foundation generalizable to diverse use cases requiring multiple environments beyond CL. We call for extending our approach into these learning paradigms to validate our claims.

\subsection{R\&D opportunities for the MDE community}

Our work highlights some research and development opportunities for the MDE community. Drawing on the experiences from developing our approach, we discuss some of these opportunities.

\subsubsection{The role of domain-specific languages}

There is a clear need for targeted DSLs at various points of this approach, making a good case for DSL engineering as a research direction in this space. While DSLs excel at raising the level of abstraction in software engineering~\cite{mernik2005when}, existing approaches to RL environment specification remain largely imperative and framework-specific (e.g., Gymnasium's API). To leverage MDE techniques for the benefit of RL and ML, DSLs are needed that can capture structural properties (e.g., states, transitions, and structural relations of environments), behavioral semantics (reward structures, termination conditions), and parametric variability for generating families of environments.

Beyond environment specification, DSLs could also capture curriculum structures~\cite{soviany2022curriculum} (e.g., sequential curricula or curriculum graphs), difficulty progression strategies, and partial constraints over environment families. This aligns with recent efforts in developing DSLs for RL~\cite{sinani2024towards} but calls for more support for family-level reasoning and search-based generation, which remain unexplored.

Of course, case-by-case DSL engineering does not scale with the high volume of RL-based applications and therefore, automated DSL engineering techniques should be investigated, e.g., based on domain ontologies~\cite{joaovarandapereira2016ontological} and the structure of the learning paradigm.

\subsubsection{HOTs for generating mutation MTs}
Our approach relies on mutation MTs that are typically developed in a manual fashion. We support simple automation for the derivation of MTs from states that are marked as mutable by the RL expert, and the value of such derivation is especially clear in highly heterogeneous state spaces in which manual MT specification is infeasible. This demonstrates the opportunity for higher-order transformations (HOTs)~\cite{tisi2009use} to generate mutation operators automatically.
HOTs could be used to derive mutation operators from the metamodel structure~\cite{khne2010explicit} (e.g., type hierarchies, multiplicities), ensure semantic preservation (e.g., solvability constraints), and encode domain heuristics (e.g., monotonic difficulty adjustments).

Similar avenues have been explored before in related problems, e.g., in generating search operators for model-driven optimization~\cite{burdusel2021automatic}. We recommend extensions of such foundations toward machine learning paradigms and the related mutation synthesis problems. In addition to generating mutations, HOTs could, e.g., generate operators that explicitly control environment difficulty gradients or diversity distributions.
In addition, combining HOTs with meta-learning signals (e.g., feedback from RL performance) opens a direction for adaptive mutation operators that evolve alongside the learning process. Such mechanisms are of high utility, e.g., in adaptive CL, in which the curriculum is dynamically adjusted based on the agent's learning progress~\cite{matiisen2019teacher}.

\subsubsection{Variability and product family engineering}
Families of RL environments can be naturally interpreted through variability modeling and product-family engineering~\cite{muthig2002model-driven}. In our approach, environments share a common structure (chiefly defined by the initial environment) and mutations induce variability---analogous to deriving products from a shared platform~\cite{lackner2017chapter}. Unlike in traditional product lines, variability, here, is continuous and generative, rather than based on selecting from predefined configurations.
Extensions of traditional variability techniques may bring benefits to our approach. In particular, variability models could be enriched with structural and parametric variation operators (beyond Boolean features), objective functions (e.g., diversity, complexity), and relations between variants (e.g., ordering in curricula).
Such extensions could bridge product-line engineering with model-driven optimization~\cite{semerth2020diversity}, enabling automated exploration of environment families rather than manual configuration.
Global decision-making over deep variability~\cite{kienzle2023global} can aid reasoning about families of learning environments rather than individual variants.
\section{Related work}\label{sec:related-work}

\textit{Product-family engineering}
is a systematic approach to create families of related products that share a common set of core assets while maintaining managed variability to address diverse requirements~\cite{lackner2017chapter}.
PFE enables reuse of core assets across variants, reducing development costs~\cite{lameh2025modeling}.
However, PFE uses a discrete selection from predefined variants.
Our approach uses continuous mutations guided by measures for environment generation, instead of enumerable feature combinations.

\textit{Procedural content generation (PCG)}
methods automatically generate diverse game content, such as maps, levels, and environments~\cite{johnson2010cellular}. 
While PCG can automatically produce game levels, these techniques are domain-specific and generate environments narrowly tailored to game engines, limiting reuse, especially in problems other than games~\cite{togelius2013procedural}.
Moreover, ensuring that generated environments meet validity constraints, such as preventing invalid or unsolvable environments, is still challenging~\cite{silva2025procedural}.

\textit{Large language model (LLM)}-based environment generation prompts LLMs with textual context to synthesize environment specifications and executable code~\cite{hu2025agentgen}.
However, LLMs provide no formal guarantees about environment validity. Their stochastic, non-deterministic nature prevents systematic verification of environment properties or structural constraints~\cite{zhang2025position}.
Furthermore, LLM-based generation requires substantial computational resources to consistently generate and select valuable environments~\cite{liang2024eurekaverse}.

\textit{Genetic algorithm}-based curriculum generation evolves environment sequences to improve training performance.  
For example, \citet{song2022robust} operate directly on simulator encodings and focus on online curriculum generation, and they extend this into multi-agent settings~\cite{song2024genetic}. 
However, these approaches lack structural validity guarantees, explicit family-level reasoning, constraint-aware generation, and modeled mutation semantics. 
Our approach instead contributes an MDE perspective, and decouples environment-family generation from training.

\textit{Model instance generation and model-based design space exploration (DSE)} are related topics to ours. Refinery~\cite{semerth2020diversity} is a framework for the automated generation of consistent and diverse graph models for rich test instances. In VIATRA DSE~\cite{abdeen2014multi-objective}, multi-objective optimization rules are captured in model transformations, enabling a smooth integration of genetic algorithms into MDE problems.
Such frameworks can be integrated into our approach to replace some of the prototype components we developed. 

\textit{MDE for RL} type works that relate to ours are the following.
\citet{gatto2019modeling} generate training and deployment code interfacing with simulators via Robot Operating System. However, their approach models the RL algorithm, treating the environment as an external system.
\citet{molderez2019marlon} propose Marlon, a DSL bridging multi-agent RL and distributed systems, where the environment is taken as given. 
\citet{sinani2024towards} propose RLML, a modeling language for specifying RL problems, but require manual enumeration of states, actions, and rewards for a single environment. 
\citet{kusmenko2022model-driven} translate game descriptions into RL environments, but their toolchain is limited to turn-based games. 
\citet{liaskos2025model-driven} generate RL training environments from goal models, but focus on a single environment.
In this work, we address these limitations by proposing an approach for developing families of RL environments.
\section{Conclusion}\label{sec:conclusion}

In this work, we presented a model-driven approach for the automated construction of families of reinforcement learning training environments. Reinforcement learning often necessitates training processes over families of environments, e.g., to train the agent on a sequence of environments with gradually increasing complexity. However, obtaining such a family of environments through manual or software code-level mutations is a labor-intensive and error-prone endeavor.
Our approach automates the generation of environments through a model-based hybrid genetic algorithm, in which environment mutation operators and constraints are codified in model transformations, and executed in a state-of-the-art incremental model transformation engine.

Our work demonstrates the utility of the model-driven engineering body of knowledge in modern machine learning problems. Accordingly, in this work, we identify opportunities for the model-driven engineering community to contribute to addressing such problems through their expertise.

Future work will focus on the performance aspects of the approach, e.g., through incrementalization and algorithm tuning, as well as evaluation on an industry-scale case.
\section*{Acknowledgement}

We acknowledge the support of the Natural Sciences and Engineering Research Council of Canada (NSERC), DGECR-2024-00293 (End-to-end Sustainable Systems Engineering).

\bibliographystyle{ACM-Reference-Format}
\bibliography{bib/references}

@article{barriga2022parmorel,
  title = {PARMOREL: a framework for customizable model repair},
  author = {Barriga, Angela and Heldal, Rogardt and Rutle, Adrian and Iovino, Ludovico},
  year = {2022},
  journal = {Soft. Sys. Mod.},
  publisher = {Springer},
  volume = {21},
  number = {5},
  pages = {1739--1762}
}

@inproceedings{bengio2009curriculum,
  title = {Curriculum learning},
  author = {Bengio, Yoshua and Louradour, J\'{e}r\^{o}me and Collobert, Ronan and Weston, Jason},
  year = {2009},
  booktitle = {Proceedings of the 26th Annual International Conference on Machine Learning},
  publisher = {ACM},
  series = {ICML '09},
  pages = {41–48},
  doi = {10.1145/1553374.1553380},
}

@inproceedings{bergmann2015viatra,
  title = {VIATRA 3: {A} Reactive Model Transformation Platform},
  author = {G{\'{a}}bor Bergmann and Istvan David and {\'{A}}bel Heged{\"{u}}s and {\'{A}}kos Horv{\'{a}}th and Istv{\'{a}}n R{\'{a}}th and Zolt{\'{a}}n Ujhelyi and D{\'{a}}niel Varr{\'{o}}},
  year = {2015},
  booktitle = {Theory and Practice of Model Transformations - 8th International Conference, ICMT\@STAF 2015, L'Aquila, Italy, July 20-21, 2015. Proceedings},
  publisher = {Springer},
  series = {LNCS},
  volume = {9152},
  pages = {101--110},
  doi = {10.1007/978-3-319-21155-8\_8},
}

@article{burdusel2021automatic,
	title        = {Automatic generation of atomic multiplicity-preserving search operators for search-based model engineering},
	author       = {Burdusel, Alexandru and Zschaler, Steffen and John, Stefan},
	year         = 2021,
	journal      = {Soft. Sys. Mod.},
	publisher    = {Springer},
	volume       = 20,
	number       = 6,
	pages        = {1857--1887}
}

@inproceedings{cobbe2020leveraging,
  title = {Leveraging Procedural Generation to Benchmark Reinforcement Learning},
  author = {Cobbe, Karl and Hesse, Chris and Hilton, Jacob and Schulman, John},
  year = {2020},
  booktitle = {Proc of the 37th International Conference on Machine Learning},
  publisher = {PMLR},
  series = {PMLR},
  volume = {119},
  pages = {2048--2056},
}

@book{cover1991elements,
  title = {Elements of Information Theory},
  author = {{T.M. Cover} and Thomas, Joy A},
  year = {1991},
  publisher = {John Wiley \& Sons},
  address = {Nashville, TN},
  series = {Wiley Series in Telecommunications and Signal Processing},
  edition = {99},
}

@book{cowell2011measuring,
  title = {Measuring Inequality},
  author = {Cowell, Frank},
  year = {2011},
  publisher = {Oxford University Press},
  address = {London, England},
  series = {London School of Economics Perspectives in Economic Analysis},
  edition = {3}
}

@inproceedings{dagenais2025complex,
  title = {Complex Model Transformations by Reinforcement Learning with Uncertain Human Guidance},
  author = {Dagenais, Kyanna and David, Istvan},
  year = {2025},
  booktitle = {2025 ACM/IEEE 28th International Conference on Model Driven Engineering Languages and Systems (MODELS)},
  doi = {10.1109/MODELS67397.2025.00025},
}

@inproceedings{david2022devs,
  title = {{DEVS} Model Construction as a Reinforcement Learning Problem},
  author = {David, Istvan and Syriani, Eugene},
  year = {2022},
  booktitle = {2022 Annual Modeling and Simulation Conference (ANNSIM)},
  pages = {30--41},
  doi = {10.23919/ANNSIM55834.2022.9859369},
  organization = {IEEE},
}

@inbook{david2024automated,
  title = {Automated Inference of Simulators in Digital Twins},
  author = {David, Istvan and Syriani, Eugene},
  year = {2024},
  booktitle = {{Handbook of Digital Twins}},
  publisher = {{CRC Press}},
  isbn = {9781032546070},
  doi = {10.1201/9781003425724-11},
  chapter = {8},
  pages = {122--148},
}

@inproceedings{dennis2020emergent,
  title = {Emergent Complexity and Zero-shot Transfer via Unsupervised Environment Design},
  author = {Dennis, Michael and Jaques, Natasha and Vinitsky, Eugene and Bayen, Alexandre and Russell, Stuart and Critch, Andrew and Levine, Sergey},
  year = {2020},
  booktitle = {Advances in Neural Information Processing Systems},
  publisher = {Curran Associates, Inc.},
  volume = {33},
  pages = {13049--13061},
}

@book{eiben2015introduction,
  title = {Introduction to evolutionary computing},
  author = {Eiben, Agoston E and Smith, James E},
  year = {2015},
  publisher = {Springer}
}

@inproceedings{eisenberg2021towards,
  title = {Towards Reinforcement Learning for In-Place Model Transformations},
  author = {Eisenberg, Martin and Pichler, Hans-Peter and Garmendia, Antonio and Wimmer, Manuel},
  year = {2021},
  booktitle = {2021 ACM/IEEE 24th Intl Conf. on Model Driven Engineering Languages and Systems (MODELS)},
  pages = {82--88}
}

@article{figueiredoprudencio2024survey,
  title = {A Survey on Offline Reinforcement Learning: Taxonomy, Review, and Open Problems},
  author = {Figueiredo Prudencio, Rafael and Maximo, Marcos R. O. A. and Colombini, Esther Luna},
  year = {2024},
  journal = {IEEE Trans Neural Netw Learn Syst},
  volume = {35},
  number = {8},
  pages = {10237--10257},
  doi = {10.1109/TNNLS.2023.3250269}
}

@inproceedings{gatto2019modeling,
  title = {Modeling Deep Reinforcement Learning Based Architectures for Cyber-Physical Systems},
  author = {Gatto, Nicola and Kusmenko, Evgeny and Rumpe, Bernhard},
  year = {2019},
  booktitle = {2019 ACM/IEEE 22nd International Conference on Model Driven Engineering Languages and Systems Companion},
  pages = {196--202},
  doi = {10.1109/MODELS-C.2019.00033}
}

@article{hospedales2022meta-learning,
  title = {Meta-Learning in Neural Networks: A Survey},
  author = {Hospedales, Timothy and others},
  year = {2022},
  journal = {IEEE Transactions on Pattern Analysis and Machine Intelligence},
  volume = {44},
  number = {9},
  pages = {5149--5169},
  doi = {10.1109/TPAMI.2021.3079209}
}

@inproceedings{john2019searching,
  title = {Searching for optimal models: Comparing two encoding approaches},
  author = {John, Stefan and Burdusel, Alexandru and Bill, Robert and Struber, Daniel and Taentzer, Gabriele and Zschaler, Steffen and Wimmer, Manuel},
  year = {2019},
  booktitle = {12th International Conference on Model Transformations ICMT 2019},
  pages = {1--22}
}

@article{john2023graph,
  title = {A graph-based framework for model-driven optimization facilitating impact analysis of mutation operator properties},
  author = {John, Stefan and Kosiol, Jens and Lambers, Leen and Taentzer, Gabriele},
  year = {2023},
  journal = {Soft. Sys. Mod.},
  publisher = {Springer},
  volume = {22},
  number = {4},
  pages = {1281--1318}
}

@inproceedings{kim2021survey,
  title = {A survey on simulation environments for reinforcement learning},
  author = {Kim, Taewoo and Jang, Minsu and Kim, Jaehong},
  year = {2021},
  booktitle = {2021 18th International Conference on Ubiquitous Robots (UR)},
  pages = {63--67},
  organization = {IEEE}
}

@inproceedings{kusmenko2022model-driven,
  title = {A Model-Driven Generative Self Play-Based Toolchain for Developing Games and Players},
  author = {Kusmenko, Evgeny and others},
  year = {2022},
  booktitle = {Proceedings of the 21st ACM SIGPLAN International Conference on Generative Programming: Concepts and Experiences},
  publisher = {ACM},
  series = {GPCE 2022},
  pages = {95–107},
  doi = {10.1145/3564719.3568687},
}

@incollection{lackner2017chapter,
  title = {Chapter Four - Advances in Testing Software Product Lines},
  author = {Hartmut Lackner and Bernd-Holger Schlingloff},
  year = {2017},
  publisher = {Elsevier},
  series = {Advances in Computers},
  volume = {107},
  pages = {157--217},
  doi = {https://doi.org/10.1016/bs.adcom.2017.07.001},
  issn = {0065-2458}
}

@article{lameh2025modeling,
  title = {Modeling variability in product line engineering (PLE) for systems engineering (SE)},
  author = {Lameh, Jos\'{e} and Dubray, Alexandra and Jankovic, Marija},
  year = {2025},
  journal = {Proceedings of the Design Society},
  volume = {5},
  pages = {2491–2500},
  doi = {10.1017/pds.2025.10263}
}

@article{liang2024eurekaverse,
  title = {Eurekaverse: Environment curriculum generation via large language models},
  author = {Liang, William and Wang, Sam and Wang, Hung-Ju and Bastani, Osbert and Jayaraman, Dinesh and Ma, Yecheng Jason},
  year = {2024},
  journal = {arXiv preprint arXiv:2411.01775}
}

@inproceedings{liaskos2025model-driven,
  title = {Model-Driven Design and~Generation of~Training Simulators for~Reinforcement Learning},
  author = {Liaskos, Sotirios and M. Khan, Shakil and Mylopoulos, John and Golipour, Reza},
  year = {2025},
  booktitle = {Conceptual Modeling},
  publisher = {Springer},
  pages = {170--191},
}

@article{liu2025ai,
  title = {AI Simulation by Digital Twins: Systematic Survey, Reference Framework, and Mapping to a Standardized Architecture},
  author = {Liu, Xiaoran and David, Istvan},
  year = {2025},
  journal = {Software and Systems Modeling},
  doi = {10.1007/s10270-025-01306-0},
}

@inproceedings{liu2026reference,
  title = {A Reference Architecture of Reinforcement Learning Frameworks},
  author = {Liu, Xiaoran and David, Istvan},
  year = {2026},
  booktitle = {2026 IEEE 23rd International Conference on Software Architecture (ICSA)},
  doi = {10.1109/ICSA66085.2026.00016},
}

@inproceedings{molderez2019marlon,
  title = {Marlon: {A} domain-specific language for multi-agent reinforcement learning on networks},
  author = {Molderez, Tim and Oeyen, Bjarno and De Roover, Coen and De Meuter, Wolfgang},
  year = {2019},
  booktitle = {Proc of the 34th ACM/SIGAPP Symposium on Applied Computing},
  publisher = {ACM},
  pages = {1322–1329},
  doi = {10.1145/3297280.3297413}
}

@inproceedings{malik2021when,
  title = {When Is Generalizable Reinforcement Learning Tractable?},
  author = {Malik, Dhruv and Li, Yuanzhi and Ravikumar, Pradeep},
  year = {2021},
  booktitle = {Advances in Neural Information Processing Systems},
  publisher = {Curran Associates, Inc.},
  volume = {34},
  pages = {8032--8045},
}

@article{marcinandrychowicz2020learning,
  title = {Learning dexterous in-hand manipulation},
  author = {OpenAI: Marcin Andrychowicz and others},
  year = {2020},
  journal = {The International Journal of Robotics Research},
  volume = {39},
  number = {1},
  pages = {3--20},
  doi = {10.1177/0278364919887447}
}

@article{narvekar2020curriculum,
  title = {Curriculum Learning for Reinforcement Learning Domains: A Framework and Survey},
  author = {Sanmit Narvekar and Bei Peng and Matteo Leonetti and Jivko Sinapov and Matthew E. Taylor and Peter Stone},
  year = {2020},
  journal = {J Machine Learning Research},
  volume = {21},
  number = {181},
  pages = {1--50},
}

@article{naveed2024model,
  title = {Model driven engineering for machine learning components: A systematic literature review},
  author = {Naveed, Hira and Arora, Chetan and Khalajzadeh, Hourieh and Grundy, John and Haggag, Omar},
  year = {2024},
  journal = {Inf Softw Technol},
  publisher = {Elsevier},
  volume = {169},
  pages = {107423}
}

@article{sinani2024towards,
  title = {Towards a Domain-Specific Modelling Environment for Reinforcement Learning},
  author = {Sinani, Natalie and others},
  year = {2024},
  journal = {arXiv preprint arXiv:2410.09368}
}

@inproceedings{sodhani2021multi-task,
  title = {Multi-task reinforcement learning with context-based representations},
  author = {Sodhani, Shagun and Zhang, Amy and Pineau, Joelle},
  year = {2021},
  booktitle = {International conference on machine learning},
  pages = {9767--9779},
  organization = {PMLR}
}

@article{soviany2022curriculum,
  title = {Curriculum Learning: A Survey},
  author = {Soviany, Petru and others},
  year = {2022},
  journal = {International Journal of Computer Vision},
  volume = {130},
  number = {6},
  pages = {1526--1565},
  doi = {10.1007/s11263-022-01611-x}
}

@book{sutton1998reinforcement,
  title = {Reinforcement learning: An introduction},
  author = {Sutton, Richard S and Barto, Andrew G},
  year = {1998},
  publisher = {MIT press Cambridge}
}

@book{thulasiraman2011graphs,
  title = {Graphs: theory and algorithms},
  author = {Thulasiraman, Krishnaiyan and Swamy, Madisetti NS},
  year = {2011},
  publisher = {John Wiley \& Sons}
}

@inproceedings{tobin2017domain,
  title = {Domain randomization for transferring deep neural networks from simulation to the real world},
  author = {Tobin, Josh and Fong, Rachel and Ray, Alex and Schneider, Jonas and Zaremba, Wojciech and Abbeel, Pieter},
  year = {2017},
  booktitle = {2017 IEEE/RSJ international conference on intelligent robots and systems (IROS)},
  pages = {23--30},
  organization = {IEEE}
}

@article{wang2022survey,
  title = {A Survey on Curriculum Learning},
  author = {Wang, Xin and others},
  year = {2022},
  journal = {IEEE Transactions on Pattern Analysis and Machine Intelligence},
  volume = {44},
  number = {9},
  pages = {4555--4576},
  doi = {10.1109/TPAMI.2021.3069908}
}

@inproceedings{yu2020meta-world,
  title = {Meta-World: A Benchmark and Evaluation for Multi-Task and Meta Reinforcement Learning},
  author = {Yu, Tianhe and others},
  year = {2020},
  booktitle = {Proceedings of the Conference on Robot Learning},
  publisher = {PMLR},
  series = {Proceedings of Machine Learning Research},
  volume = {100},
  pages = {1094--1100},
}

@article{zhang2022survey,
  title = {A Survey on Multi-Task Learning},
  author = {Zhang, Yu and Yang, Qiang},
  year = {2022},
  journal = {IEEE Transactions on Knowledge and Data Engineering},
  volume = {34},
  number = {12},
  pages = {5586--5609},
  doi = {10.1109/TKDE.2021.3070203}
}

@inproceedings{zhao2020sim-to-real,
  title = {Sim-to-Real Transfer in Deep Reinforcement Learning for Robotics: a Survey},
  author = {Zhao, Wenshuai and others},
  year = {2020},
  booktitle = {2020 IEEE Symposium Series on Computational Intelligence (SSCI)},
  pages = {737--744},
  doi = {10.1109/SSCI47803.2020.9308468}
}

@misc{liu2023towards,
  title = {Towards Out-Of-Distribution Generalization: A Survey},
  author = {Jiashuo Liu and Zheyan Shen and Yue He and Xingxuan Zhang and Renzhe Xu and Han Yu and Peng Cui},
  year = {2023},
  url = {https://arxiv.org/abs/2108.13624},
  eprint = {2108.13624},
  archiveprefix = {arXiv},
  primaryclass = {cs.LG}
}

@inproceedings{pitkevich2024survey,
  title = {A Survey on Sim-to-Real Transfer Methods for Robotic Manipulation},
  author = {Pitkevich, Andrei and Makarov, Ilya},
  year = {2024},
  booktitle = {IEEE Intl Symposium on Intelligent Systems and Informatics (SISY)},
  pages = {000259--000266},
  doi = {10.1109/SISY62279.2024.10737545}
}

@inproceedings{wang2025rgdr,
  title={RGDR: Reward-Guided Domain Randomization for Autonomous Driving},
  author={Wang, Dejin and Ghoreishi, Seyede Fatemeh},
  booktitle={2025 IEEE 28th International Conference on Intelligent Transportation Systems (ITSC 2025), IEEE},
  year={2025}
}

@article{puterman1990markov,
  title = {Markov decision processes},
  author = {Puterman, Martin L},
  year = {1990},
  journal = {Handbooks in operations research and management science},
  publisher = {Elsevier},
  volume = {2},
  pages = {331--434}
}

@inproceedings{johnson2010cellular,
  title={Cellular automata for real-time generation of infinite cave levels},
  author={Johnson, Lawrence and Yannakakis, Georgios N and Togelius, Julian},
  booktitle={Proceedings of the 2010 Workshop on Procedural Content Generation in Games},
  pages={1--4},
  year={2010}
}

@misc{togelius2013procedural,
  title={Procedural content generation: Goals, challenges and actionable steps},
  author={Togelius, Julian and Champandard, Alex J and Lanzi, Pier Luca and Mateas, Michael and Paiva, Ana and Preuss, Mike and Stanley, Kenneth O},
  year={2013},
  publisher={Schloss Dagstuhl--Leibniz-Zentrum fuer Informatik}
}

@article{silva2025procedural,
  title={Procedural game level generation with GANs: potential, weaknesses, and unresolved challenges in the literature},
  author={Silva, Daniele F and Torchelsen, Rafael P and Aguiar, Marilton S},
  journal={Multimedia Tools and Applications},
  pages={1--27},
  year={2025},
  publisher={Springer}
}

@inproceedings{hu2025agentgen,
  title = {AgentGen: Enhancing Planning Abilities for Large Language Model based Agent via Environment and Task Generation},
  author = {Hu, Mengkang and others},
  year = {2025},
  booktitle = {Proceedings of the 31st ACM SIGKDD Conference on Knowledge Discovery and Data Mining V.1},
  publisher = {ACM},
  series = {KDD '25},
  pages = {496–507},
  doi = {10.1145/3690624.3709321}
}

@inproceedings{zhang2025position,
  title = {Position: Trustworthy {AI} Agents Require the Integration of Large Language Models and Formal Methods},
  author = {Yedi Zhang and others},
  year = {2025},
  booktitle = {Forty-second International Conference on Machine Learning Position Paper Track},
  url = {https://openreview.net/forum?id=wkisIZbntD}
}

@inproceedings{ntentos2024supporting,
  title = {Supporting architectural decision making on training strategies in reinforcement learning architectures},
  author = {Ntentos, Evangelos and Warnett, Stephen John and Zdun, Uwe},
  year = {2024},
  booktitle = {21st Intl Conf on Software Architecture (ICSA)},
  pages = {90--100},
  organization = {IEEE}
}

@inproceedings{song2022robust,
  title={Robust reinforcement learning via genetic curriculum},
  author={Song, Yeeho and Schneider, Jeff},
  booktitle={2022 International Conference on Robotics and Automation (ICRA)},
  pages={5560--5566},
  year={2022},
  organization={IEEE}
}

@misc{towers2025gymnasium,
  title = {Gymnasium: A Standard Interface for Reinforcement Learning Environments},
  author = {Mark Towers and others},
  year = {2025},
  doi = {10.48550/arXiv.2407.17032}
}

@inproceedings{song2024genetic,
  title = {Genetic Algorithm for Curriculum Design in Multi-Agent Reinforcement Learning},
  author = {Yeeho Song and Jeff Schneider},
  year = {2024},
  booktitle = {8th Annual Conference on Robot Learning},
  url = {https://openreview.net/forum?id=2CScZqkUPZ}
}

@article{watkins1992q-learning,
  title = {Q-learning},
  author = {Watkins, Christopher JCH and Dayan, Peter},
  year = {1992},
  journal = {Machine learning},
  publisher = {Springer},
  volume = {8},
  number = {3},
  pages = {279--292}
}

@article{matiisen2019teacher,
  title={Teacher--student curriculum learning},
  author={Matiisen, Tambet and Oliver, Avital and Cohen, Taco and Schulman, John},
  journal={IEEE transactions on neural networks and learning systems},
  volume={31},
  number={9},
  pages={3732--3740},
  year={2019},
  publisher={IEEE}
}

@inproceedings{narvekar2016source,
  title={Source task creation for curriculum learning},
  author={Narvekar, Sanmit and Sinapov, Jivko and Leonetti, Matteo and Stone, Peter},
  booktitle={Proceedings of the 2016 international conference on autonomous agents \& multiagent systems},
  pages={566--574},
  year={2016}
}

@article{bertsimas1993simulated,
  title = {{Simulated Annealing}},
  author = {Dimitris Bertsimas and John Tsitsiklis},
  year = {1993},
  journal = {Statistical Science},
  publisher = {Institute of Mathematical Statistics},
  volume = {8},
  number = {1},
  pages = {10 -- 15},
  doi = {10.1214/ss/1177011077}
}

@article{el2006hybrid,
  title={Hybrid Genetic Algorithms: A Review.},
  author={El-Mihoub, Tarek A and Hopgood, Adrian A and Nolle, Lars and Battersby, Alan}
}

@article{moscato1989evolution,
  title={On evolution, search, optimization, genetic algorithms and martial arts: Towards memetic algorithms},
  author={Moscato, Pablo and others}
}

@article{shannon1948mathematical,
  title = {A mathematical theory of communication},
  author = {Shannon, C. E.},
  year = {1948},
  journal = {The Bell System Technical Journal},
  volume = {27},
  number = {3},
  pages = {379--423},
  doi = {10.1002/j.1538-7305.1948.tb01338.x}
}

@inproceedings{florensa2017reverse,
  title = {Reverse Curriculum Generation for Reinforcement Learning},
  author = {Florensa, Carlos and Held, David and Wulfmeier, Markus and Zhang, Michael and Abbeel, Pieter},
  year = {2017},
  booktitle = {Proceedings of the 1st Annual Conference on Robot Learning},
  publisher = {PMLR},
  series = {Proceedings of Machine Learning Research},
  volume = {78},
  pages = {482--495},
}

@inproceedings{tisi2009use,
  title = {On the Use of Higher-Order Model Transformations},
  author = {Tisi, Massimo and Jouault, Fr\'{e}d\'{e}ric and Fraternali, Piero and Ceri, Stefano and B\'{e}zivin, Jean},
  year = {2009},
  booktitle = {Proceedings of the 5th European Conference on Model Driven Architecture - Foundations and Applications},
  publisher = {Springer},
  series = {ECMDA-FA '09},
  pages = {18–33},
  doi = {10.1007/978-3-642-02674-4_3},
  numpages = {16}
}

@article{joaovarandapereira2016ontological,
  title = {Ontological approach for DSL development},
  author = {Pereira, Maria Joao Varanda and Fonseca, Joao and Henriques, Pedro Rangel},
  year = {2016},
  journal = {Computer Languages, Systems \& Structures},
  volume = {45},
  pages = {35--52},
}

@inproceedings{muthig2002model-driven,
  title = {Model-Driven Product Line Architectures},
  author = {Muthig, Dirk and Atkinson, Colin},
  year = {2002},
  booktitle = {Software Product Lines},
  publisher = {Springer},
  pages = {110--129},
  isbn = {978-3-540-45652-0},
}

@inproceedings{kienzle2023global,
  title = {Global Decision Making Over Deep Variability in Feedback-Driven Software Development},
  author = {Kienzle, Joerg and others},
  year = {2023},
  booktitle = {Proceedings of the 37th IEEE/ACM International Conference on Automated Software Engineering},
  publisher = {ACM},
  series = {ASE '22},
  doi = {10.1145/3551349.3559551},
  articleno = {178},
  numpages = {6},
}

@inproceedings{elaasar2023opencaesar,
  title = {openCAESAR: Balancing Agility and Rigor in Model-Based Systems Engineering},
  author = {Elaasar, Maged and Rouquette, Nicolas and Wagner, David and Oakes, Bentley James and Hamou-Lhadj, Abdelwahab and Hamdaqa, Mohammad},
  year = {2023},
  booktitle = {2023 ACM/IEEE International Conference on Model Driven Engineering Languages and Systems Companion (MODELS-C)},
  pages = {221--230},
  doi = {10.1109/MODELS-C59198.2023.00051},
}

@article{semerth2020diversity,
  title = {Diversity of graph models and graph generators in mutation testing},
  author = {Semer{\'a}th, Oszk{\'a}r and Farkas, Rebeka and Bergmann, G{\'a}bor and Varr{\'o}, D{\'a}niel},
  year = {2020},
  journal = {Int J Softw Tools Technol Transf},
  volume = {22},
  number = {1},
  pages = {57--78},
  doi = {10.1007/s10009-019-00530-6},
  issn = {1433-2787}
}

@inproceedings{abdeen2014multi-objective,
  title = {Multi-objective optimization in rule-based design space exploration},
  author = {Abdeen, Hani and Varr\'{o}, D\'{a}niel and Sahraoui, Houari and Nagy, Andr\'{a}s Szabolcs and Debreceni, Csaba and Heged\"{u}s, \'{A}bel and Horv\'{a}th, \'{A}kos},
  year = {2014},
  booktitle = {Proceedings of the 29th ACM/IEEE International Conference on Automated Software Engineering},
  publisher = {ACM},
  series = {ASE '14},
  pages = {289–300},
  doi = {10.1145/2642937.2643005},
}

@inproceedings{terry2021pettingzoo,
  title = {{P}etting{Z}oo: {G}ym for {M}ulti-{A}gent {R}einforcement {L}earning},
  author = {Terry, Jordan and others},
  year = {2021},
  booktitle = {Advances in Neural Information Processing Systems},
  publisher = {Curran Associates, Inc.},
  volume = {34},
  pages = {15032--15043}
}

@misc{makoviychuk2021isaac,
  title = {{I}saac {G}ym: {H}igh {P}erformance {GPU}-{B}ased {P}hysics {S}imulation {F}or {R}obot {L}earning},
  author = {Viktor Makoviychuk and others},
  year = {2021},
  doi = {10.48550/arXiv.2108.10470}
}

@article{mernik2005when,
  title = {When and how to develop domain-specific languages},
  author = {Mernik, Marjan and Heering, Jan and Sloane, Anthony M.},
  year = {2005},
  month = dec,
  journal = {ACM Comput. Surv.},
  publisher = {ACM},
  volume = {37},
  number = {4},
  pages = {316–344},
  doi = {10.1145/1118890.1118892},
}

@book{ross2022simulation,
  title = {Simulation},
  author = {Ross, Sheldon M},
  year = {2022},
  publisher = {academic press}
}

@book{russell2020artificial,
  title = {Artificial intelligence},
  author = {Russell, Stuart and Norvig, Peter},
  year = {2020},
  publisher = {Pearson},
  address = {Upper Saddle River, NJ},
  edition = {4},
}

@inproceedings{khne2010explicit,
  title = {Explicit Transformation Modeling},
  author = {K{\"u}hne, Thomas and Mezei, Gergely and Syriani, Eugene and Vangheluwe, Hans and Wimmer, Manuel},
  year = {2010},
  booktitle = {Models in Software Engineering},
  publisher = {Springer},
  pages = {240--255},
  isbn = {978-3-642-12261-3},
}

@inproceedings{shephard1974law,
  title = {The Law of Diminishing Returns},
  author = {Shephard, Ronald W. and F{\"a}re, Rolf},
  year = {1974},
  booktitle = {Production Theory},
  publisher = {Springer},
  pages = {287--318},
  doi = {10.1007/978-3-642-80864-7_17}
}

@inproceedings{tsoy2003influence,
  title = {The influence of population size and search time limit on genetic algorithm},
  author = {Tsoy, Y.R.},
  year = {2003},
  booktitle = {7th Korea-Russia International Symposium on Science and Technology, Proceedings KORUS 2003. (IEEE Cat. No.03EX737)},
  volume = {3},
  pages = {181--187 vol.3}
}

@inproceedings{mclean2025meta-world+,
  title = {Meta-World+: An Improved, Standardized, {RL} Benchmark},
  author = {Reginald McLean and Evangelos Chatzaroulas and Luc McCutcheon and Frank R{\"o}der and Tianhe Yu and Zhanpeng He and K.R. Zentner and Ryan Julian and J K Terry and Isaac Woungang and Nariman Farsad and Pablo Samuel Castro},
  year = {2025},
  booktitle = {The Thirty-ninth Annual Conference on Neural Information Processing Systems Datasets and Benchmarks Track},
}

@inproceedings{chaffre2020sim-to-real,
  title = {{Sim-to-Real Transfer with Incremental Environment Complexity for Reinforcement Learning of Depth-Based Robot Navigation}},
  author = {Chaffre, Thomas and Moras, Julien and Chan-Hon-Tong, Adrien and Marzat, Julien},
  year = {2020},
  booktitle = {{Proceedings of the 17th International Conference on Informatics, Automation and Robotics, ICINCO 2020}},
  pages = {314--323},
  url = {https://ensta.hal.science/hal-02958155},
}

\end{document}